\documentclass{aa}       % New LaTeX A&A Standard Fonts
\usepackage{graphicx}
\usepackage{amssymb}
\usepackage{amsmath}
\usepackage{epsfig}

\begin{document}

\title{The role of Kelvin-Helmholtz instability in the
internal structure of relativistic outflows. The case of the jet
in 3C 273}

\author{
M. Perucho\inst{1,2} \and A.P. Lobanov\inst{2},
J.-M.$^{\underline{a}}$ Mart\'{\i}\inst{1} \and P.E.
Hardee\inst{3} }

\authorrunning{Perucho et al.}
\titlerunning{The role of K-H instability in relativistic outflows}

\institute{Departament d'Astronomia i Astrof\'{\i}sica,
Universitat de Val\`encia, C/Dr. Moliner 50, 46100 Burjassot,
Spain \and Max-Planck-Institut f\"ur Radioastronomie, Auf dem
H\"ugel 69, 53121 Bonn, Germany \and Department of Physics and
Astronomy, University of Alabama, Tuscaloosa, AL 35487  }

\offprints{M. Perucho, \email{perucho@mpifr-bonn.mpg.de}}

\date{Received <date> / Accepted <date>}

\abstract {Relativistic outflows represent one of the best-suited
tools to probe the physics of AGN.  Numerical modelling of
internal structure of the relativistic outflows on parsec scales
provides important clues about the conditions and dynamics of the
material in the immediate vicinity of the central black holes in
AGN.} {We investigate possible causes of the structural patterns
and regularities observed in the parsec-scale jet of the
well-known quasar 3C 273.} {We present here the results from a 3D
relativistic hydrodynamics numerical simulation based on the
parameters given for the jet by Lobanov \& Zensus (2001), and one
in which the effects of jet precession and the injection of
discrete components have been taken into account. We compare the
model with the structures observed in 3C 273 using very long
baseline interferometry and constrain the basic properties of the
flow.} {We find growing perturbation modes in the simulation with
similar wavelengths to those observed, but with a different set of
wave speeds and mode identification. If the observed longest
helical structure is produced by the precession of the flow,
longer precession periods should be expected.} {Our results show
that some of the observed structures could be explained by growing
Kelvin-Helmholtz instabilities in a slow moving region of the jet.
However, we point towards possible errors in the mode
identification that show the need of more complete linear analysis
in order to interpret the observations. We conclude that, with the
given viewing angle, superluminal components and jet precession
cannot explain the observed structures.}
%, thereby alleviating the constraints on the
%kinetic power and energy content of the jet.} {}


\maketitle

\section{Introduction}
\label{sect1}

The structure and kinematics of parsec-scale outflows is typically
explained in terms of shocks (Marscher \cite{mar80}, Marscher \&
Gear \cite{mg85}, G\'omez et al. \cite{gam93}, \cite{gam94},
G\'omez et al. \cite{gom+94}) and Kelvin-Helmholtz (K-H)
instabilities (Hardee \cite{har82}, \cite{har84}, \cite{har87},
Hardee et al. \cite{hcr97}, Hardee \cite{har00}, \cite{har03},
Hardee et al. \cite{har05}) developing in a relativistic fluid.
Relativistic shocks may dominate the jet dynamics and emission at
small scales, but are likely to dissipate at distances larger than
$\sim 10$\,pc (Lobanov \& Zensus \cite{lz99}) due to the
interaction with the slower flow. On intermediate scales ($\sim
10$--100\,pc) shocks and plasma instabilities may play equally
important roles in jets (Lobanov \& Roland \cite{lr01}).
Distributions of the synchrotron turnover frequency obtained for
3C\,273 (Lobanov et al. 1997) and 3C\,345 (Lobanov 1998) indicate
that both shocks and instabilities are present on these scales,
while larger scales are most likely dominated by plasma
instabilities alone.

Recent studies by Hardee (\cite{har00}) have shown that K-H
instability may produce complex, three-dimensional ribbon-like and
thread-like patterns inside a relativistic jet. In these ribbons
and threads, a substantial increase of particle pressure and radio
emissivity can be expected. This model has been successfully
applied to the jet in 3C 120 (Hardee \cite{har03}, Hardee et al.
\cite{har05}). The threaded structure forming a double helix has
been detected in a space VLBI image of 3C\,273 made at 5\,GHz
(Lobanov et al. \cite{lob+00}). It was explained in terms of K-H
instability developing in a relativistic flow with a modest
Lorentz factor $\gamma= 2.1$ and a relativistic Mach number $M =
3.5$ (Lobanov \& Zensus 2001, hereafter \cite{LZ01}). The
analytical approach used in LZ01 is based on linear perturbation
analysis of a K-H instability developed by Hardee (\cite{har87},
\cite{har00}). A similar approach applied to kiloparsec-scale jet
in M87 allowed for accurate determination of physical parameters
and modelling of radio emission to be made (Lobanov et al.
\cite{lhe03}).

However, results from numerical simulations of relativistic flows
indicate that, after the linear regime of instability growth, the
jets can be easily disrupted (Perucho et al. \cite{pe+04b}). In
addition to this, the bulk Lorentz factor $\gamma=2.1$ derived in
LZ01 is below the values required to explain the apparent speeds
of $\beta_\mathrm{app}\sim 5-8\,c$ of enhanced emission features
observed in 3C\,273. LZ01 suggest that the K-H instability is
developing in a slower, underlying flow, and the fast components
are most likely faster shock waves produced in the jet by the
periodic ejections associated with the nuclear flares. The
presence of such shocks may disrupt the linear growth of the K-H
instability as well, and it is not clear whether the linear
stability analysis can be still applied in the presence of these
kind of non-linear effects in the flow.

In view of these concerns, it is important to confront the results
of LZ01 with numerical simulations, and attempt to address several
fundamental issues about the stability and propagation of
relativistic flows similar to the one observed in 3C\,273.
Numerical simulations can be used to verify whether the linear
theory can be applied for explaining self-consistently the
morphology and kinematics of parsec-scale flows, and whether these
flows preserve fingerprints of linear modes even when the
non-linear regime has developed or non-linear features, such as
fast components, appear. Numerical simulations provide a means to
address these problems by following the transition from linear to
non-linear regimes of instability development (Perucho et al.
\cite{pe+04a,pe+04b}). The ultimate goal of this work is to probe
the advantages and limitations of the combination of different
approaches (linear theory, numerical simulations and observations)
to studies of parsec-scale jets.

3C\,273 is the second quasar discovered (Hazard et al.
\cite{hms63}), and the first one for which the emission lines were
identified with red-shifted hydrogen lines (Schmidt \cite{sch63}).
In the same work, Schmidt (\cite{sch63}) also pointed out the
presence of a jet-like structure in this object. During the last
four decades, the active nucleus and the relativistic outflow in
3C\,273 have been studied in great detail (Courvoisier
\cite{cou98}).  The parsec-scale radio jet in 3C\,273 has been
monitored for almost three decades (Pearson et al. \cite{pea+81},
Unwin et al. \cite{unw+85}, \cite{unw+89}, Zensus et al.
\cite{zen+88}, \cite{zen+90}, Davis et al. \cite{dum91}, Abraham
et al. \cite{abr+96}, Krichbaum et al. \cite{kwz00}, Lobanov et
al. \cite{lob+00}, Asada et al. \cite{asa+02}). The emission
associated with the relativistic outflow on kiloparsec scales has
been probed extensively in the radio (Conway et al. \cite{con+81},
\cite{con+93}), near infrared (Neumann et al. \cite{nmr97},
Hutchings et al. \cite{hut+04}), optical (Thompson et al.
\cite{tmw93}, Jester \cite{jes01}, Jester et al. \cite{jes+01})
and X-ray (R\"oser et al. \cite{roe+00}, Marshall et al.
\cite{mar+01}, Sambruna et al. \cite{sam+01}) wavebands.

The relativistic jet observed in the quasar 3C273 is one-sided,
with no signs of emission on the counter-jet side at dynamic
ranges of up to 16,000:1 (Unwin et al.  \cite{unw+85}).  This is
evidence for strong relativistic boosting in an intrinsically
double-sided outflow powered by an accretion disk around the
central black hole (Begelman et al. \cite{bbr84}). The mass of the
central black hole in 3C\,273 is estimated to be $M_\mathrm{bh} =
5.5^{+0.9}_{-0.8}\times 10^8\mathrm{M}_{\sun}$ (Kaspi et al.
\cite{kas+00}).  The enhanced emission features (jet components)
identified in the jet on scales of up to $\sim$20 milliarcseconds
(mas) are moving at apparent speeds exceeding the speed of light
by factors of 5-8 (Abraham et al. \cite{abr+96}). Plausible ranges
of the Lorentz factor $\gamma\approx 5$--10 and viewing angles
$\theta_\mathrm{jet}\approx 10\degr$--$15\degr$ have been inferred
from these measurements.

Ejections of new components into the jet occur roughly once every
year (Krichbaum et al. \cite{kwz00}), and they are likely to be
related to weak optical flares observed with a similar periodicity
(Belokon \cite{bel81}).  The position angle at which the
components are ejected shows regular variations with a likely
period of about 13--15 years (Abraham et al. \cite{abr+96},
Abraham \& Romero \cite{ar99}), correlated with the long-term
variability observed in 3C\,273 in the optical (Babadzhanyants \&
Belokon \cite{bb93}) and radio (Turler et al. \cite{tcp99}) bands.
Abraham \& Romero (\cite{ar99}) have suggested that this
periodicity may reflect changes of the jet axis induced by the
relativistic precession of the inner part of the accretion disk.

Results from the linear analysis and numerical modelling are
presented, compared and discussed in Sect.~\ref{sect3} in
connection to explaining the observed properties of parsec-scale
outflow in 3C\,273. Main results of the work are discussed in
Sect.~\ref{sect4}.

Throughout the paper, we adopt the flat $\Lambda$CDM Cosmology
with the Hubble constant $H_{0}=71\,h\,$km\, s$^{-1}$\,Mpc$^{-1}$,
where $h$ is a constant with a likely value of 1, and matter
density $\Omega_{M}=0.27$. The positive definition of spectral
index, $S\propto\nu^{\alpha}$ is used. For 3C\,273 ($z=0.157$,
Strauss et al. \cite{str+92}), the adopted cosmological parameters
correspond to the luminosity distance $D_{\mathrm
L}=0.7h^{-1}$\,Gpc. The respective linear scale is
$2.69h^{-1}$\,pc\,mas$^{-1}$, and a proper motion of 1\,mas/yr
corresponds to an apparent speed of $10.1 h^{-1}\,c$.

\section{Numerical simulations}
\label{sect3}

In this section, we present numerical simulations performed with
the aim to provide a counterpart to the analytical modelling of
the jet structure in 3C\,273 made in LZ01. To give a better
account of the connection between the analytical and numerical
approaches, basic results from the linear model of LZ01 are
briefly summarized below.

\subsection{Results from the linear analysis}
\label{sect2}

Linear perturbation analysis of Kelvin-Helmholtz instability (cf.
Hardee \cite{har87,har00} and Hardee et al. \cite{hcr97}) was
applied in LZ01 to explaining the internal structure of the jet in
3C\,273. The locations of two thread-like features identified
inside the jet (the features P1 and P2, in the nomenclature of
LZ01) were approximated by combinations of several oscillatory
modes. These modes were identified with different modes of
Kelvin-Helmholtz instability. The parameters of these modes are
given in Table~\ref{tab:3c2731}. The characteristic wavelengths of
different instability modes can be related to the jet speed
($\beta_j$), Mach number ($M_j$) and the density ratio
($\eta=\rho_j/\rho_a$) between the jet and the ambient medium.
These wavelengths are obtained by approximating the dispersion
relation in the small frequency limit ($\omega\rightarrow 0$) and
large frequency limit ($\omega\gg$). Both are obtained for highly
supersonic jets ($M_j\gg 1$). The longest unstable wavelength,
$\lambda^l_{nm}$ (obtained in the low-frequency limit), and the
resonant wavelength, $\lambda^*_{nm}$ (obtained in the high
frequency limit), are given by (Hardee \cite{har87}):

\begin{equation}\label{long}
\lambda^l_{nm}=\frac{4 \gamma_j R_j (M_j^2-1)^{1/2}}{n+2m-1/2}\, ,
\end{equation}
\begin{eqnarray}\label{maxim}
\lambda^*_{nm}=\frac{2\pi}{\beta_{s,a}/\beta_j\,
(n+m+1/2)} \nonumber\\
\qquad \qquad \qquad
\frac{\gamma_j(M_j^2-\beta_j^2)^{1/2}}{(M_a^2-\beta_j^2)^{1/2}+\gamma_j(M_j^2-\beta_j^2)^{1/2}}\,
,
\end{eqnarray}
where $\beta_{s,a}$ is the sound speed of the external medium in
units of the speed of light, $\gamma_j$ is the Lorentz factor of
the jet, $M_a=\beta_j/\beta_{s,a}$, $R_j$ is the jet radius, and
$n$ (the azimuthal number) and $m$  give the number of nodes
around the jet surface and the number of nodes between the axis
and the surface, respectively. The first equation gives the
longest unstable wavelength for a body mode ($m>0$) and the zero
frequency limit for a surface mode ($m=0$), as the latter do not
show the long wavelength cut, and the second stands for the most
unstable wavelength of a given mode. A total of five wavelengths
were identified from fits to the double ridge line found in the
observations presented in LZ01. Physical parameters of the jet
were obtained in LZ01 by relating these wavelengths to the
wavelengths of the oscillatory modes from the fit to the internal
structure of the jet. The jet parameters obtained are given in
Table~\ref{tab:3c2732}. Four of the observed wavelengths were
associated with resonant wavelengths of helical and elliptical
modes and the longest wavelength (18 mas) was associated with the
helical surface mode driven externally at approximately twice the
resonant wavelength.

\begin{table}
\caption{Identified wavelengths, modes and their amplitudes from
observations (LZ01). $H$ stands for helical, $E$ for elliptical
modes, and subscripts refer to surface ($s$, fundamental) or body
modes ($b$, reflection). The latter are followed by the index
identifying the exact body mode.$^*$ stands for identified
resonant modes.} \label{tab:3c2731} \centering
\begin{tabular}{ccc}
\hline\hline
$\lambda$ [mas]&Amplitude [mas]&Mode\\
\hline
P1  P2& P1  P2&\\
18&1.5&$H_s$\\
12&1.4&$E^*_s$\\
3.9  4.1&2.2  1.5&$H^*_{b1}$\\
3.8&1.2&$E^*_{b1}$\\
1.9&0.25&$H^*_{b2}$\\
\hline
\end{tabular}
\end{table}

\begin{table*}
\caption{Jet parameters from the fit. $\gamma_j$ is Lorentz
factor, $M_{j,r}$ is the relativistic Mach number, $\eta$ is the
jet-to-ambient rest mass density ratio, $\phi_j$ is jet half
opening angle, $\theta_j$ is jet viewing angle, $c_{s,j,a}$ are
sound speeds, and $l$ is the projected linear
scale.}\label{tab:3c2732} \centering
\begin{tabular}{ccccccccc}
\hline\hline
$\gamma_j$&$M_{j,r}$&$\eta$&$R_j$[pc]&$\phi_j$ [$^\circ$]&$\theta_j$ [$^\circ$]&
$c_{s,j}$ [$c$]&$c_{s,a}$ [$c$]&$l$ [$\rm{pc/mas}$]\\
\hline
2.1&3.5&0.023&0.8&1.5&15&0.53&0.08&2.43\\
\hline
\end{tabular}
\end{table*}

It should be noted that the Lorentz factor $\gamma=2.1$ derived
for the jet is below the values given by other authors in order to
explain superluminal motions observed in 3C\,273. This may result
from Kelvin-Helmholtz instability developing in an underlying,
slower flow, and not in the flow that contains ballistic,
superluminal features (LZ01).

The results from the linear stability analysis are compared to the
numerical solutions of the stability problem obtained from the
individual simulation runs described below.

\subsection{General properties of the numerical models}
%\subsection{Setup}
Numerical simulations were performed using a three-dimensional
finite-difference code based on a high-resolution shock-capturing
scheme which solves the equations of relativistic hydrodynamics
written in conservation form. This code is an upgrade to 3D of the
code described in Mart\'{\i} et al. (\cite{m97}) and shares many
features with the 3D code GENESIS (Aloy et al. \cite{a99}). It was
parallelised using OMP directives. Simulations were performed in
an SGI Altix 3000.

In the numerical models, the initial properties of a stationary
flow in pressure equilibrium with the external medium are set
according to the results of the linear modelling (see
Table~\ref{tab:3c2732}). It has to be noted, however, that no
opening angle has been taken into account. Perturbations are
applied at the inlet. Boundary conditions are: 1)~injection at the
inlet (with the parameters given by LZ01), and 2)~outflow at the
side boundaries and at the axial end of the grid. An extended grid
with a decreasing resolution is added on each side of the main
grid and at its axial end, in order to avoid spurious numerical
reflections of the solution at the main grid boundaries.

In order to achieve a steady initial model, we add a smooth
transition (i.e., shear layer) between the jet and the ambient
medium, of the form:
\begin{equation}\label{shearlayer1}
\rho (r)=\rho_a - (\rho_a-\rho_j)/cosh(r)^m,
\end{equation}
\begin{equation}\label{shearlayer2}
v_z (r) = v_{z,0}/cosh(r)^m,
\end{equation}
where $\rho$ stands for rest mass density and $v_z$ for axial
velocity ($v_{z,0}$ is the value at the axis corresponding to the
Lorentz factor $\gamma_j=2.1$), subscripts $a$ and $j$ correspond
to ambient medium an jet, respectively, and $r$ is the radial
coordinate. The smaller the resolution, the smaller the exponent
$m$ has to be in order to reduce the numerical noise below the
amplitudes of the perturbations.

We performed two simulations to investigate the general
development of a K-H instability in the flow and to analyse the
effect of fast components and the jet precession. In the first
simulation (3C273-A) we perturb a stationary flow so as to observe
which modes and wavelengths dominate the jet structure. In the
second simulation (3C273-B) we try to see if similar instability
structures can be generated by the precession of the jet and
periodic injections of faster components into the flow. In both
simulations, the parameters of the steady flow are those from LZ01
(see Table~\ref{tab:3c2732}). We use the perfect gas equation of
state with the adiabatic exponent $\Gamma=4/3$.

\subsection{Simulation 3C273-A}
\subsubsection{Initial setup}
In this simulation we introduce perturbations at frequencies
calculated such that they are expected to reproduce the observed
wavelengths in the jet structure if these are propagating at the
given speed in LZ01. The grid extends over 844\,\rm{cells} in the
axial direction and 128\,\rm{cells} in both lateral directions
(including the extended grid). We use a resolution of $16\,
\rm{cells}/R_j$, with $R_j$ the radius of the jet, in the
transversal direction and $4\,\rm{cells}/R_j$ in the direction of
the flow. Simulation lasted for a time $1097\, R_j/c$ (i.e., about
$2852 \,\rm{yrs}$ when scaled to source units; see next
paragraph), and it used $\simeq 11$ Gb of RAM memory and $8$
processors during around $30$ days in a SGI Altix 3000 computer.

\begin{figure*}[!t]
\centerline{
  \includegraphics[width=0.8\textwidth]{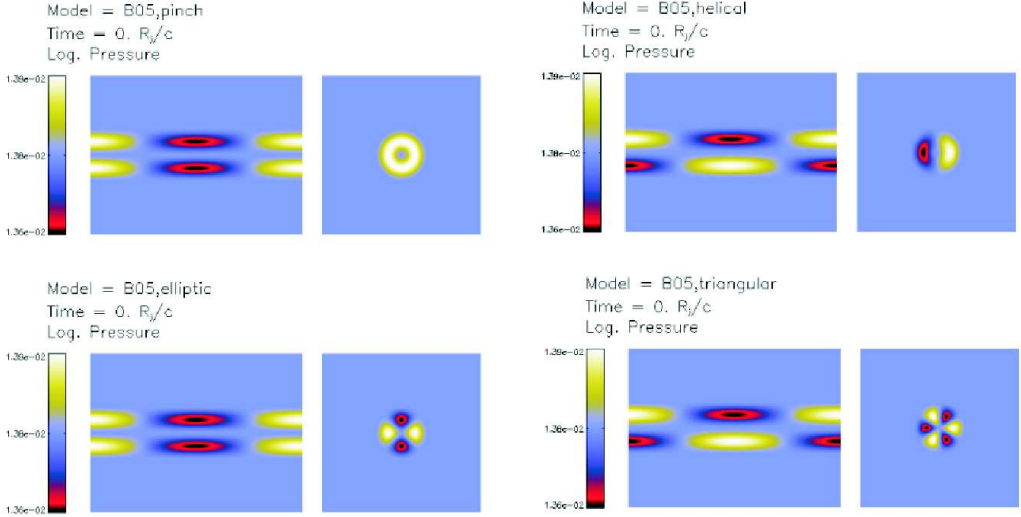}}
\caption{Structure of perturbations (axial and transversal cuts)
in a generic 3D jet, as described by Eq. (\ref{3Dpert}). Top left:
pinching mode ($n=0$). Top right: helical mode ($n=1$). Bottom
left: elliptical mode ($n=2$). Bottom right: triangular mode
($n=3$).}\label{fig:3Dperts}
\end{figure*}

Assuming an angle to the line of sight of $15^\circ$ and redshift
$z=0.158$ ($1\,\rm{mas}=2.43$ pc) the observed jet is
$169\,\rm{pc}$ long. Considering the jet radius given in LZ01
($0.8\, \rm{pc}$), the numerical grid extends over $211\, R_j$
(axial) times $8\,R_j$ times $8\,R_j$ (transversal), i.e.,
$169\,\rm{pc}\times 6.4\,\rm{pc} \times 6.4\,\rm{pc}$. This allows
us to accommodate all relevant relativistic and sub-relativistic
structures which could give rise to the wavelengths observed in
the patterns P1 and P2 from LZ01. A shear layer of $2\, R_j$ width
($m=2$) in Eqs.~(\ref{shearlayer1})--(\ref{shearlayer2}) is
included in the initial rest mass density and axial velocity
profiles to keep numerical stability of the initial jet. To avoid
reflection of the numerical noise from the main grid boundaries,
an extended grid is introduced, with a cell size increasing
progressively by 20\%, as we do not need fine cells in this
extended region because we are mainly interested in the linear
regime, which does not involve large radial distortions of the
jet. The extended grid has 24 cells in the radial directions,
reaching up to $36\,R_j$ on each side of the jet, and 168 cells
axially, reaching up to $316.5\,R_j$. Elliptical and helical modes
are induced at the inlet using the following expression:
\begin{equation}\label{3Dpert}
    P' = \frac{A_0}{\cosh^2 r} \, \cos(\omega
    t+n \, \theta)\, \sin^2(\pi r) ,
\end{equation}
where $A_0(=10^{-4})$ is the initial amplitude, $r$ is the radial
coordinate, $\omega$ is the frequency of the mode, $n=1$ for
helical modes and $n=2$ for elliptical ones, $\theta$ is the polar
angle in cylindrical coordinates, and $\sin^2(\pi r)$ is used in
order to give an initial transversal structure to the modes. The
evolution of perturbations and their coupling to K-H modes have
been shown to be independent from this transversal structure
(Perucho et al. \cite{pe+05}). In Fig.~\ref{fig:3Dperts}, we show
axial and transversal cuts for a three-dimensional jet with
periodic boundary conditions in the axial directions, like those
used in Perucho et al. (\cite{pe+05}). The axial structure shown
in the axial cuts is added to Eq.~(\ref{3Dpert}) as a term
$k_z\,z$ in the cosinus function. Typical structures induced by
Eq.~(\ref{3Dpert}) in such a generic jet are those shown in the
transversal cuts. The sum of all the input modes gives the total
perturbation. The simulation has to reproduce the resonant
wavelengths of the basic modes identified in
Table~\ref{tab:3c2732} (2, 4 and 12 mas). It should be noted that
the helical surface mode at $\lambda_\mathrm{Hs}=18$ mas is driven
externally and therefore cannot be reproduced in this simulation.

\begin{table*}
\caption{Correspondence between the observed wavelength and the
perturbations used for the simulation 3C273-A. $\lambda^{obs}$ is
the observed, projected wavelength, $\lambda^{theor}$ is the
intrinsic wavelength, and $\omega$ is the derived frequency for
the given intrinsic wavelength and wave speed (see text). Fourth
to seventh columns give the wavelengths and wave speeds derived
from the dispersion relation solution, marked in
Fig.~\ref{fig:drs} as $\omega_1$, $\omega_2$ and $\omega_3$.}
\label{tab:tw} \centering
\begin{tabular}{ccccccc}
\hline \hline $\lambda^{obs}\,[\rm{mas}]$ &
$\lambda^{theor}\,[R_j]$ & $\omega\,[c/R_j]$& Pinch b1 & Helical s & Elliptical s & Elliptical b1\\
&&&$\lambda$; $v_\omega$&$\lambda$; $v_\omega$&$\lambda$; $v_\omega$&$\lambda$; $v_\omega$\\
\hline
12 ($E^*_s$) & 110.0   & 0.013 ($\omega_1$)&-&38.5; 0.081&18.5; 0.077&-\\
4  ($H^*_{b1}, E^*_{b1}$)& 37.4    & 0.039 ($\omega_2$)&-&16.0; 0.10&6.1; 0.078&-\\
2  ($H^*_{b2}$)& 18.7    & 0.078 ($\omega_3$)&7.0; 0.088 &12.1; 0.15&5.4; 0.14&3.2; 0.08\\
\hline
\end{tabular}
\end{table*}

Frequencies of the excited modes, both helical and elliptical, are
introduced in Eq.~\ref{3Dpert}. These frequencies are computed
from the observed wavelengths in LZ01, $\lambda^{obs}$, corrected
for projection effects and relativistic motion of the wave (time
delay), with velocity $v_w$($=0.23\,c$). We use
$\omega=2\,\pi\,v_w/\lambda^{theor}$ (see Table \ref{tab:tw}),
where
\begin{equation}\label{lamb}
\lambda^{theor}=\frac{\lambda^{obs}(1-v_w/c\,\cos\theta_j)}{\sin\theta_j},
\end{equation}
where $\theta_j$ is the angle to the line of sight. We observe
that, when computing the numerical solutions for the stability
problem for the proposed jet parameters, the wavelengths and
speeds obtained for the relevant modes (see Table \ref{tab:tw})
are different to those given in LZ01. This fact may be caused by
errors in the angle to the line of sight or in the derived
parameters in LZ01, including the wave speeds which are used in
the transformation.

\subsubsection{Results}
\begin{figure*}[!t]
\centerline{
  \includegraphics[width=0.8\textwidth]{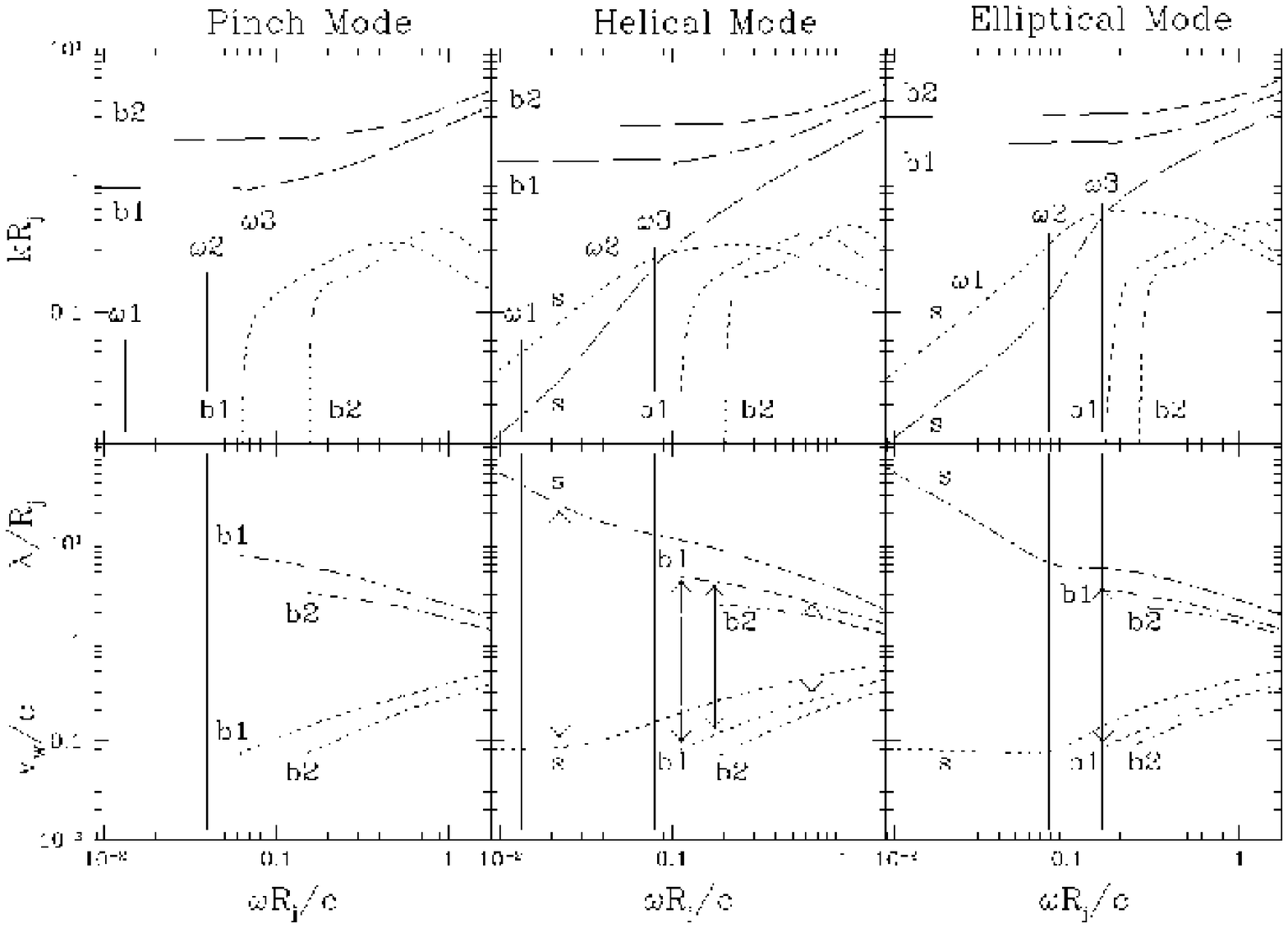}}
\caption{Solution of the stability problem for the parameters
given in Table \ref{tab:3c2732}. In the upper plots, real (dashed
lines) and imaginary (dotted lines) parts of the wavenumber are
given as a function of frequency. In the lower plots,
corresponding wavelengths and wave speeds are shown. Frequencies
$\omega1$, $\omega2$ and $\omega3$ correspond to those given in
Table \ref{tab:tw} for the simulation 3C273-A. Note that
wavelengths for the elliptical modes in the plot are for a
$180^\circ$ rotation of the ellipse and needs to be doubled for a
complete rotation. This is because such a $180^\circ$ rotation
generates a repetition of the structure in elliptical modes,
therefore giving the impression of a complete wavelength. Arrows
connecting the wavelength and wave speed plots indicate identified
modes in the simulation 3C273-A (see text).}\label{fig:drs}
\end{figure*}

To compare the results of the simulation with the numerical
solution of the stability problem, we have solved the dispersion
relation equation (e.g., Hardee \cite{har00}) for the parameters
given in Table~\ref{tab:3c2732}. Fig.~\ref{fig:drs} shows the
solutions for the pinching, helical and elliptical surface, first
and second body modes. Table~\ref{tab:drsh} shows the
characteristic wavelengths of the relevant modes.

\begin{table}
\caption{Solutions of the stability problem in the fastest growing
or resonant frequencies (indicated with an asterisk, column one
lists the frequencies, column two the wavelengths and column three
the wave speeds) and longest unstable wavelength (fourth column,
$\lambda_l$)}\label{tab:drsh} \centering
\begin{tabular}{ccccc}
\hline\hline
Mode&$\omega^*$ ($c/R_j$)&$\lambda^*$ ($R_j$)&$v_\omega^*$ ($c$)&$\lambda_l$ ($R_j$)\\
\hline
Pinch b1&0.46&3.5&0.26&7.5\\
Pinch b2&0.86&1.8&0.25&3.1\\
\hline
Helical s&0.24&7.6&0.28&-\\
Helical b1&0.66&2.5&0.26&4.5\\
Helical b2&1.07&1.5&0.25&2.4\\
\hline
Elliptic s&0.19&5.3&0.16&-\\
Elliptic b1&0.85&1.9&0.26&3.2\\
Elliptic b2&1.27&1.3&0.25&2.0\\
 \hline
\end{tabular}
\end{table}

\begin{figure*}[!t]
   \centering
   \resizebox{\hsize}{!}{\includegraphics[bb= 0 0 252 40]{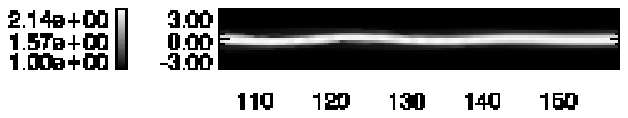}}
   \resizebox{\hsize}{!}{\includegraphics[bb= 0 0 252 40]{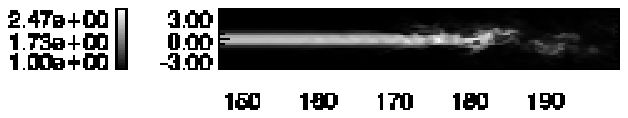}}
   \caption{Map of Lorentz factor distribution of a portion of the jet at a time before
   disruption, where a large amplitude wave is apparent (top panel, $t=320\, R_j/c$) and at
   the last frame (bottom panel, $t=1097\, R_j/c$). Coordinates are
   in jet radii. The vertical scale size is increased by a factor of 4 to better represent
   the jet structure.}
     \label{map1}
\end{figure*}

\begin{figure*}
   \centering
   \resizebox{\hsize}{!}{\includegraphics[bb= 205 295 400 450]{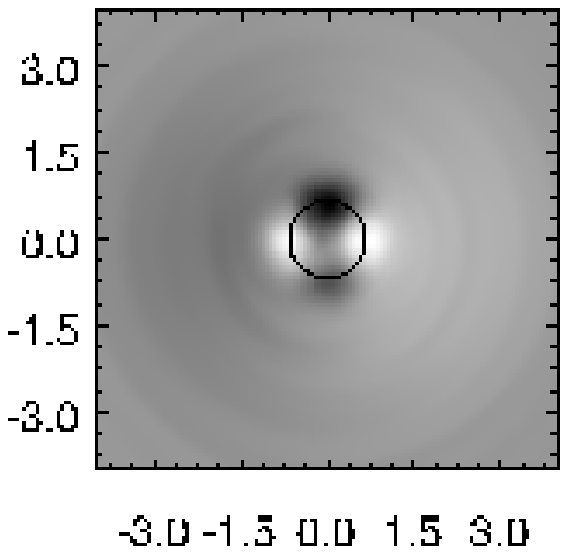}
   \includegraphics[bb= 205 295 400 450]{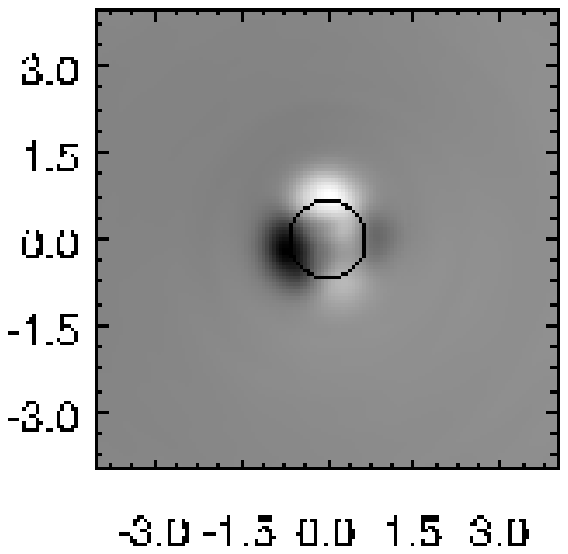}
   \includegraphics[bb= 205 295 400 450]{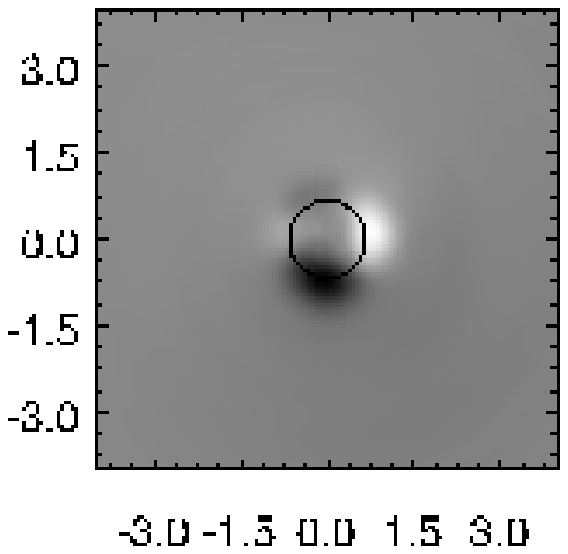}%}

%  \resizebox{\hsize}{!}{
   \includegraphics[bb= 205 295 400 450]{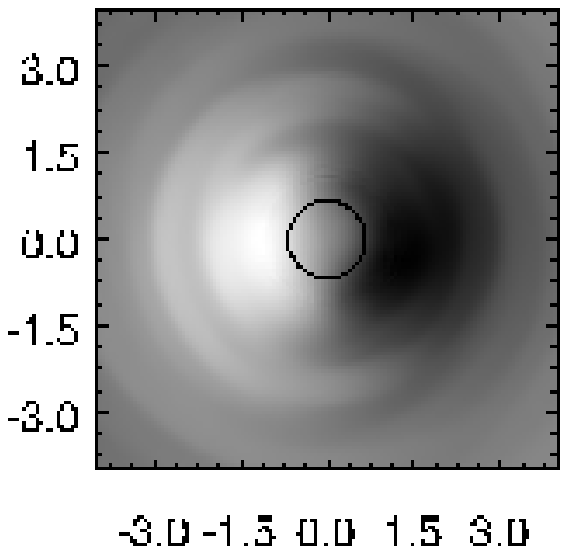}
   \includegraphics[bb= 205 295 400 450]{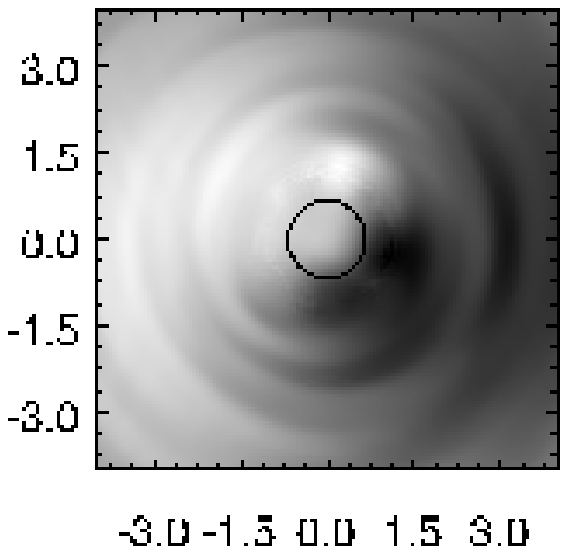}
   \includegraphics[bb= 205 295 400 450]{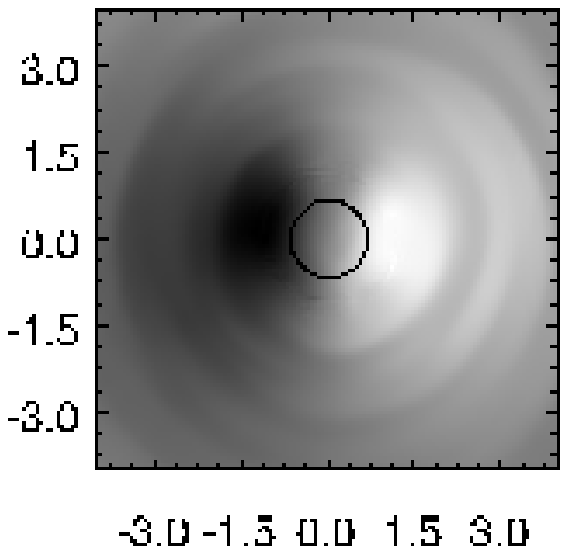}}
   \caption{Transversal structure of pressure perturbations (lighter shading indicates higher
   pressure values). Solid line indicates $v_z=0.8\,c$ contour.
   Three left panels: Cuts at $35\,R_j$, for $t=70,\,140,\,200\,R_j/c$
   where the elliptical or double helical mode rotation is apparent.
   Three right panels: Cuts at $105\,R_j$, for $t=210,\,220,\,240\,R_j/c$
   where the helical mode rotation is apparent.}
     \label{map2}
\end{figure*}

\begin{figure}[!t]
   \centering
   \resizebox{0.8\hsize}{!}{\includegraphics{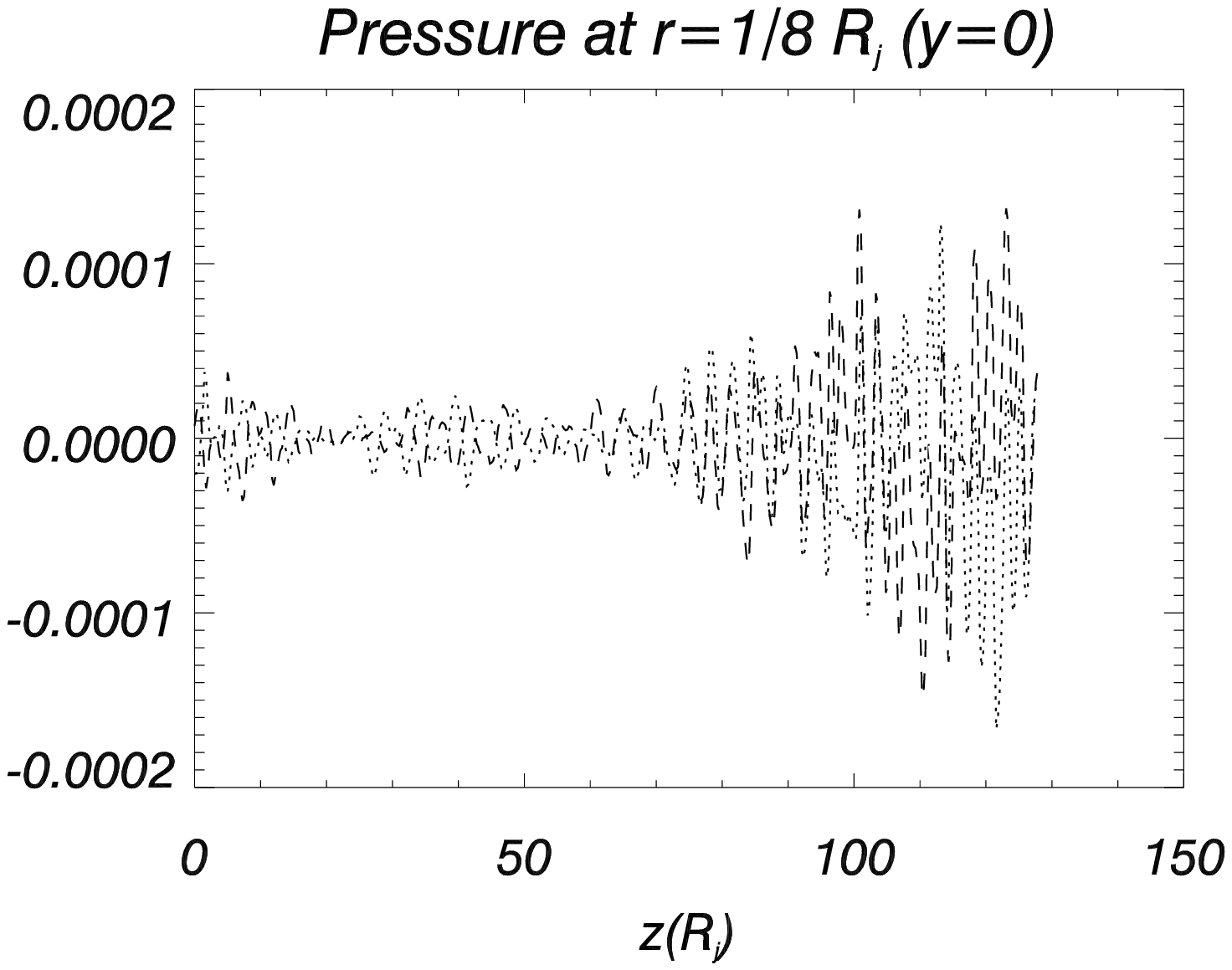} \includegraphics{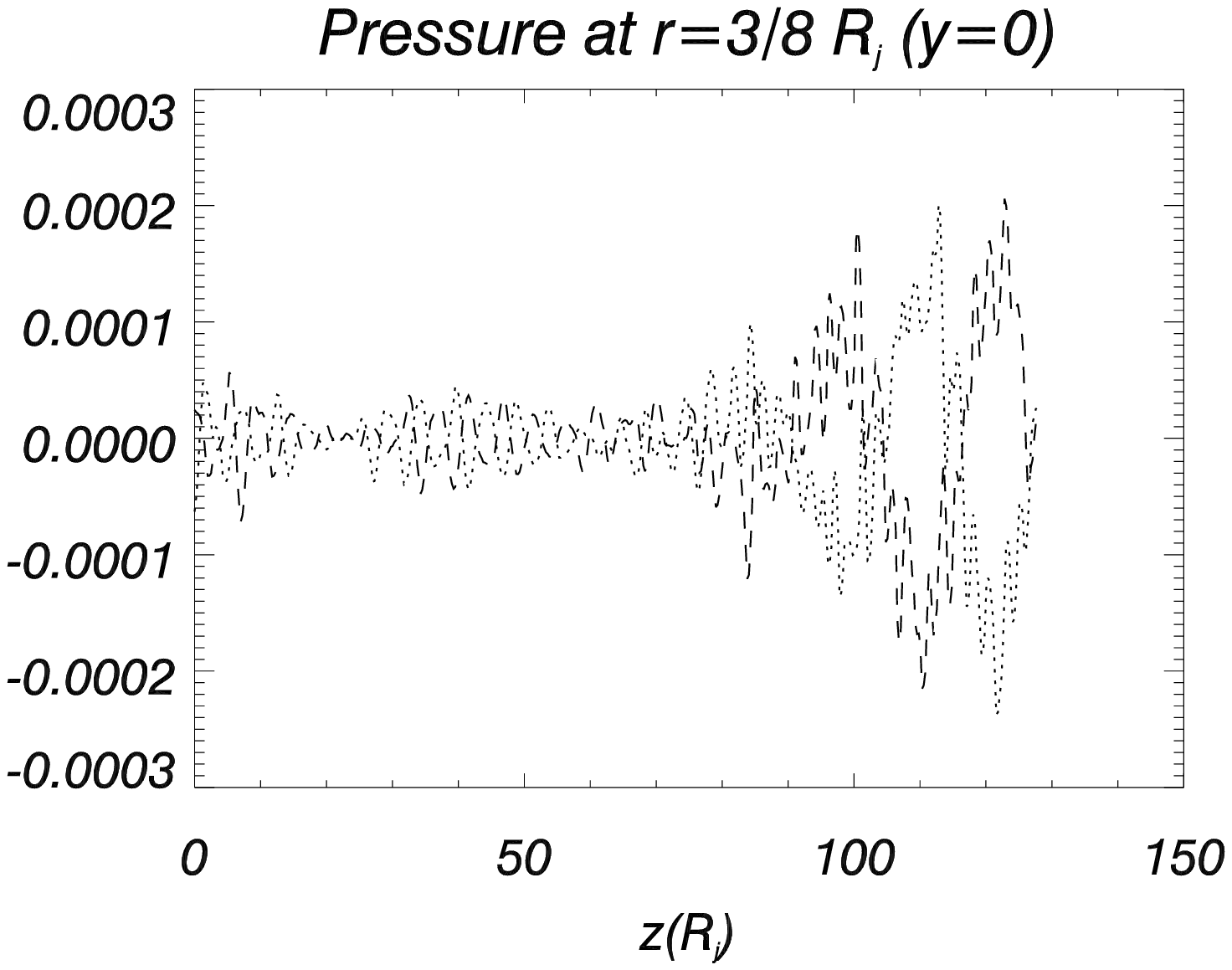}}
   \quad \resizebox{0.8\hsize}{!}{\includegraphics{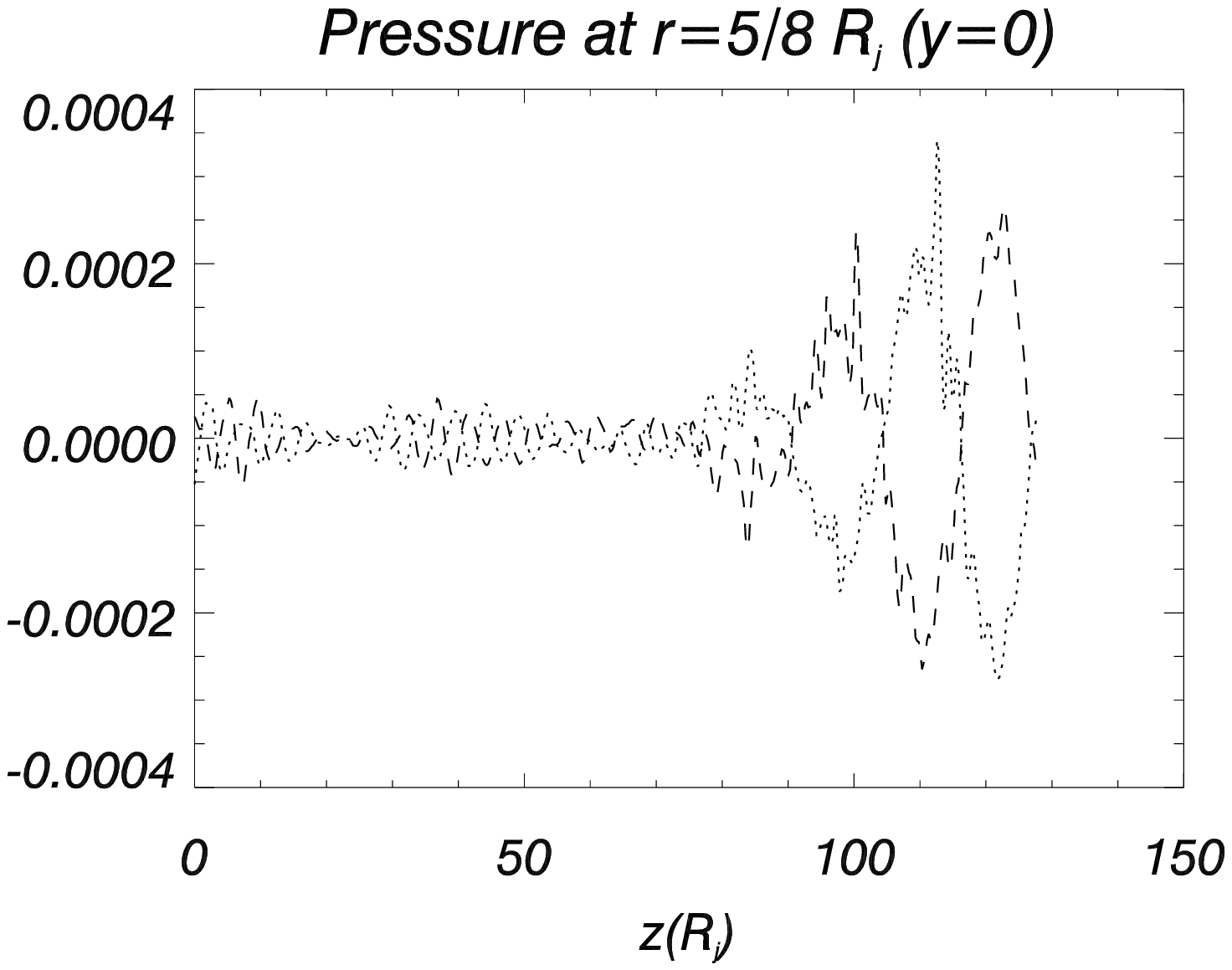} \includegraphics{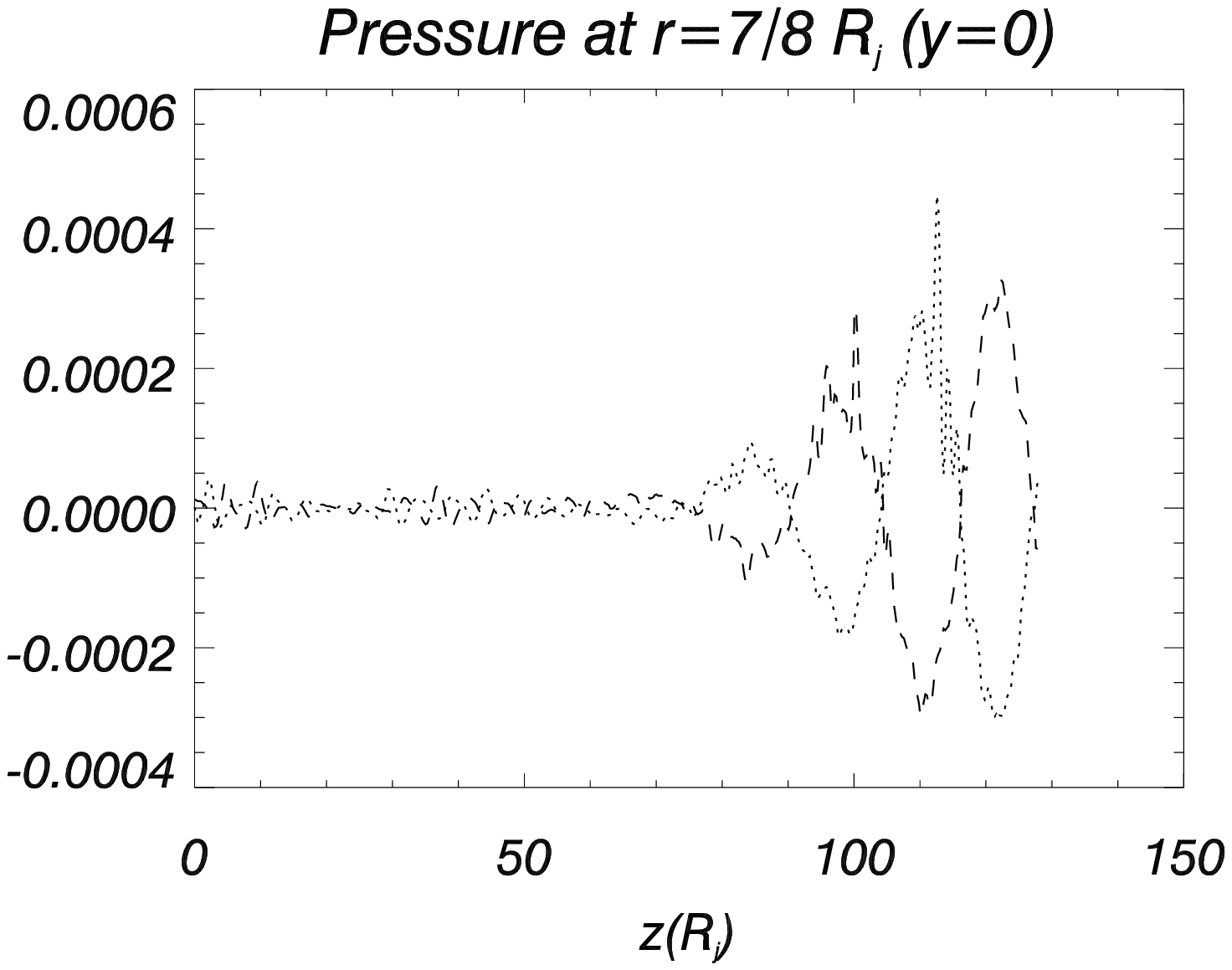}}
   \caption{Longitudinal cuts of pressure perturbation at different radii
   ($R_j/8$ top left, $3R_j/8$ top right, $5R_j/8$ bottom left,
   $7R_j/8$ bottom right) and in symmetric positions
   with respect to the jet axis at $t=250\,R_j/c$.}
     \label{pres3c273a}
\end{figure}

Fig.~\ref{map1} presents axial cuts made at two different times of
the simulation. In Fig.~\ref{map1} we observe a $\lambda\sim 20-25
R_j$ helical structure in the upper plot that could be associated
to $\omega_2$ (see Table~\ref{tab:tw}). Fig.~\ref{map2} shows
several transversal cuts of the jet illustrating competition
between the helical and elliptical modes. We can see how excited
modes dominate at different positions and times in the jet. It is
remarkable that elliptical structures show up close to the
injection point, while helical modes, develop in the jet farther
downstream. This agrees with the conclusions presented in LZ01.
Nevertheless, we have not been able to clearly identify the
elliptical mode in the longitudinal cuts. Fig.~\ref{pres3c273a}
shows longitudinal cuts of pressure perturbation (defined as the
difference between the value of the pressure in a cell and the
initial equilibrium pressure, $P-P_0$, with
$P_0\sim0.03\,\rho_a\,c^2$) at different jet radii, from which the
dominant wavelengths could be identified in the simulated jet. We
identify a $\lambda\sim 40-50\,R_j$ structure at $z<60\,R_j$ which
we interpret as due to beating between two wavelengths of the
first body helical mode at wavelengths $4.5\,R_j$ and $4\,R_j$,
like that derived from the fits to the observations by LZ01, and
given in Table~\ref{tab:3c2731} (3.9 and 4.1 mas). From plots of
pressure perturbation at different radii (Fig.~\ref{pres3c273a})
we conclude that the radial structure of this beat can only be
produced by the helical first body mode, as the fluctuations are
stronger at $R_j/2$. The beating could also be produced by the
elliptical surface mode, but the fact that pressure fluctuations
are smaller at the jet surface rules out this possibility (see,
e.g., Hardee \cite{har00}). At larger distance ($z>70 R_j$) we
have identified a large amplitude helical $\sim25\,R_j$ wavelength
and a shorter $2.5\,R_j$ wavelength superposed on the former one.
In an axial cut of the pressure perturbation close to the axis we
have identified an elliptical fluctuation with wavelength $\sim
3.5\,R_j$ at $z\sim 70-95\,R_j$ and a helical one with wavelength
$2.5\,R_j$ at $z>100\,R_j$. All the modes that are reported in
this paragraph are pointed in Fig.~\ref{fig:drs} with arrows
indicating their wavelengths and wave speeds. We should keep in
mind that the stability problem has been solved for the vortex
sheet case and the jet in the simulations has a thick shear layer.
This fact can introduce inaccuracies in the detection of modes in
the simulation.

\begin{figure}[!t]
   \centering
   \resizebox{0.8\hsize}{!}{\includegraphics{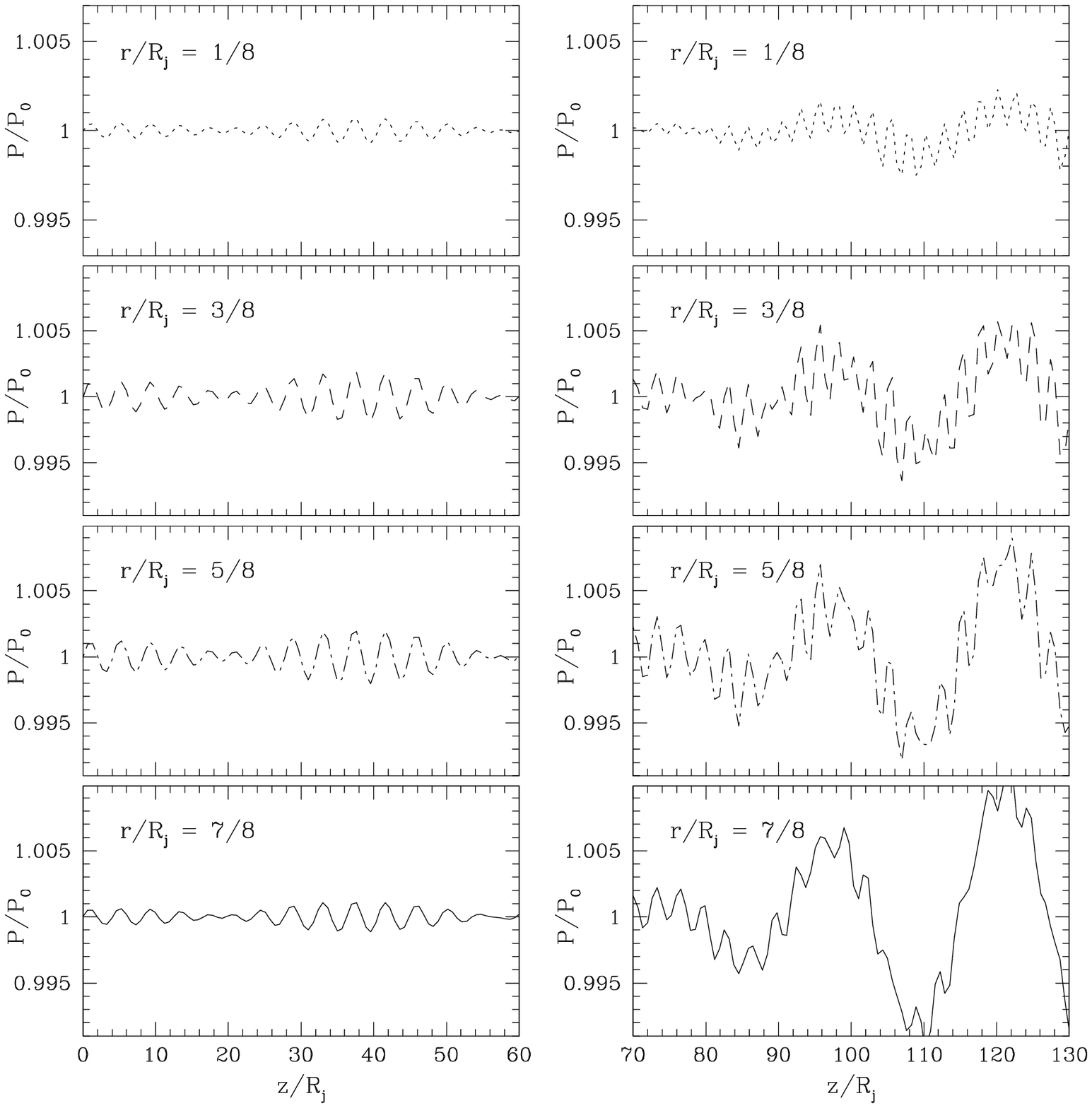}}
   \caption{Theoretical computation of the pressure perturbation produced
    by the combination of modes identified in the simulation, compared to
    Fig.~\ref{pres3c273a}. Lines indicate
    cuts at different distances from the axis. Dotted line stands
    for pressure perturbation at $R_j/8$, dashed line at $3R_j/8$,
    dash-dot line at $5R_j/8$, and long dashed line for $7R_j/8$.
    The wavelengths and wave speeds of the modes applied in this plot are
    indicated with arrows in Fig.~\ref{fig:drs}. These are the helical first
    body mode at $\lambda=4.53\,R_j$, $\lambda=4.04\,R_j$ for
    $z<60\,R_j$, the helical surface mode at $\lambda=24.3\,R_j$
    and first body mode at $\lambda=2.47\,R_j$ and the elliptical first body
    mode at $\lambda=3.22\,R_j$ for $z>70\,R_j$.
   We have used a linearly growing amplitude between $z=70\,R_j$ and
   $z=120\,R_j$ for the helical surface mode, between $z=70\,R_j$ and
   $z=105\,R_j$ for the helical first body mode, and constant amplitude from
   those distances. We have also used constant amplitude at $z<70\,R_j$ for the
   elliptical body mode.}
     \label{hardeecuts}
\end{figure}

We have used all the identified structures in order to produce
theoretical cuts of the pressure perturbation. Results are shown
in Fig.~\ref{hardeecuts}. Note that here we have plotted $P/P_0$,
whereas in Fig.~\ref{pres3c273a} we plot $P-P_0$. The similarity
to that obtained from the simulation is remarkable taking into
account the fact that the presence of a shear layer, which may
modify the stability problem solution (Perucho et al.
\cite{pe+05}), has not been considered in the interpretation of
the results.

%
%                                          Two column figure
%----------------------------------------------------------- S_vib
%
%______________________________________________________________

%
%                                                Two column figure
%----------------------------------------------------------- S_vib
%
%______________________________________________________________

%
%                                                Two column figure
%----------------------------------------------------------- S_vib

%                                                Two column figure
%----------------------------------------------------------- S_vib

When comparing the results given in the previous paragraphs with
those from LZ01, we find differences in the typical wave speeds
for the modes observed in the simulation ($\sim 0.1c$), obtained
from the solution to the stability problem (Fig.~\ref{fig:drs}),
compared to those given in LZ01 from the linear approximations
($0.23c$). This could be caused by the uncertainties of the linear
approximations, Eqs.~(\ref{long}) and (\ref{maxim}), used in LZ01.
In deriving approximations, a large classical Mach number is
assumed ($M_j=v_j/c_{s,j}\gg 1$), but this is not generally the
case for hot jets, for which $c_{s,j}\simeq 0.57$, and thus,
$M_j\le 1.75$. To investigate the uncertainties introduced by this
fact, we have used numerical solutions of the dispersion relation
for different cases and compared them with the results of the
approximations. Results show that the errors in the determination
of characteristic wavelengths with linear approximations can reach
a factor two for small Mach numbers, while for $M_j>5$, the errors
are smaller than $30\%$. This could result in significant errors
with the identification of the modes. Another difference we find
comes from the fact that we do not observe the elliptical surface
mode in the simulation, although it is fitted from observations in
LZ01 and we find in the solution to the stability problem that it
has a short growth length (Fig.~\ref{fig:drs}). This could be due
to the radial structure of the initial perturbation
($\propto\sin(\pi r)/\cosh(r)$) giving zero initial amplitudes at
the jet surfaces, and therefore suppressing the surface modes.
Otherwise we would expect the elliptical surface mode to dominate
at $z<60\,R_j$ and maybe also farther downstream, as seen in the
growth rates shown. Somehow, however, the helical surface mode
seems to develop at distances $z>70\,R_j$, and we think that this
may be due to slight changes in the radial structure of the
perturbations with downstream evolution as they grow in amplitude
and modes interact among them.

In the frame of the comparison between results from the simulation
and linear analysis and from the fits in LZ01, we now focus our
discussion on the three main structures observed in the
simulation. We define $\lambda^{sim}_1=4\,R_j$,
$\lambda^{sim}_2=25\,R_j$ and $\lambda^{sim}_3=50\,R_j$ as the
characteristic wavelengths in the simulation. Propagation speeds
of the perturbations can be measured in the section of the jet
dominated by the linear growth of instability. Although this is
difficult due to the sparsity of the data frames, we derive the
wave speed of the disruptive mode ($\lambda^{sim}_2$) following
the motion of the large amplitude wave (see Fig.~\ref{map1}) from
frame to frame, which gives $v_{w,2}\simeq 0.38\, \rm{c}$, or any
fraction of this number. From Fig.~\ref{fig:drs} we can tell that
the mode must be moving at $v_w\sim 0.09c$. Another wave speed of
the system is obtained from the rotation of the elliptical
patterns similar to those shown in Fig.~\ref{map2}, which yields
$v_{w}\simeq 0.2\,\rm{c}$. Unfortunately, we have not been able to
find the wavelength of this elliptical mode in the pressure
perturbation plots. We note that this speed is close to the value
of $v_w$ given in LZ01 and it is different from the one predicted
by the solution of the stability problem (Fig.~\ref{fig:drs}) for
the helical first body modes that we claimed to generate this long
scale structure as a beating pattern. This can be due to the
presence of a thick shear layer that certainly changes the picture
of the solution to the stability problem. It could also be due to
the structure that we have used to measure the speed having been
artificially generated by the interaction of helical modes in
$180^\circ$ phase, and therefore giving a different velocity to
those of single modes. Also, we use, as a limit for small
wavelength perturbations, a wave speed equal to that of the flow
($v_{w,1}\simeq 0.88\,c$).

The parameters of the simulated and resulting observed structures
are given in Table~\ref{tab:t0}. To reconcile the simulations with
the observational results we calculate $\lambda^{obs}$ from
$\lambda^{sim}$ ($\lambda^{theor}$ in Eq.~\ref{lamb}), using the
three different values of $v_w$ mentioned in the previous
paragraph. It should be noted that the two longest modes
identified in the simulation have an observational counterpart
when we take $v_w=0.09c$, which is a value close to that given by
the stability problem for most of the identified modes in the
simulation (see Fig.~\ref{fig:drs}). The $\lambda^{sim}_2$ mode
could be identified with the fitted helical second body mode with
the wavelength of 2 mas in LZ01. The $\lambda^{sim}_3$ mode could
correspond to the fitted first body modes at the wavelength of 4
mas in LZ01. However, $\lambda^{sim}_2$ is identified as a helical
surface mode, thus not coincident with the fitted second body mode
to the 2 mas structure and we have used the long envelope of the
beating structure in order to derive a 4 mas structure, which is
interpreted as two helical body modes with a similar wavelength in
LZ01. Using the given wave speed of $v_w=0.09c$, the term
$\sin\theta_j$ has larger influence on the result than the term
$1-v_w\cos\theta_j$ in Eq.~\ref{lamb}, so only larger angles to
the line of sight would give larger $\lambda^{obs}$ from a given
$\lambda^{sim}$: the $4\,R_j$ wavelength would result in a 1 mas
observed structure at $45^\circ$ and the $25\,R_j$ wavelength
would result in a 6 mas observed structure at the same angle to
the line of sight. Nevertheless, this angles are ruled out by jet
to counter-jet (still not observed) flux ratios and by recent
observations by Jorstad et al. (\cite{jor+05}). At that wave
speed, shorter angles to the line of sight would result in even
shorter observed wavelengths. This translates into the need of
larger wavelengths in the simulation in order to fit them to
observations, but Fig.~\ref{fig:drs} tells us that we are in the
longest unstable wavelength limit for body modes, so this seems
unrealistic.

Why we do not see the 12 mas elliptical surface mode is thought to
be due to the radial structure of the initial perturbation
($\propto\sin(\pi r)/\cosh(r)$) which, as stated above, gives zero
initial amplitudes at the jet surfaces, therefore suppressing
these modes. Moreover, the 12 mas mode, with the wave speed given
in LZ01, would require a $110\,R_j$ wavelength in the simulation,
which is difficult to observe even in a grid as large as was used
here, in particular when shorter harmonics grow fast and disrupt
the flow.

%______________________________________________________________
\begin{table*}
\caption{First column give identified wavelengths in LZ01 from
shorter to longer, in the second column we have written
characteristic wavelengths in the simulation also in increasing
order, and the last three columns give the wavelengths as observed
depending on the wave speed.} \label{tab:t0} \centering
\begin{tabular}{c|c|ccc}
\hline \hline

$\lambda^{obs}\,[\rm{mas}]$& $\lambda^{sim}\,[R_j]$&
$\lambda^{sim}_{v_w=0.09\,c}\,[\rm{mas}]$ &
$\lambda^{sim}_{v_w=0.23\,c}\,[\rm{mas}]$  &
$\lambda^{sim}_{v_w=0.88\,c}\,[\rm{mas}]$\\
 \hline
2 &  4 & 0.36 & 0.44 & 2.27\\
4 & 25 & 2.28 & 2.7  & 14.3\\
12& 50 & 4.56 & 5.5  & 28.5\\
\hline
\end{tabular}
\end{table*}

\subsubsection{Nonlinear regime}

Nonlinear effects become important at time $t=350\,R_j/c$ with the
disruption of the head of the jet due to the longest helical mode
$\lambda^{sim}_2\simeq 25\,R_j$. After that point perturbations
produced at the disruption point propagate backwards slowly as a
backflow. The disruption point itself moves downstream due to
constant injection of momentum at the inlet and the change of
conditions around the jet. The disruption point advances from
$160\, R_j$ to $180\,R_j$ by the end of the simulation (see
Fig.~\ref{map1}).

Morphology of the jet at the end of the run (lower panel of
Fig.~\ref{map1}) is thus different from the observed source mainly
due to the disruption of the jet. These difference may result from
the development of a disruptive mode in the simulation which is
not present in the real jet due to, for example, magnetic fields
or an opening angle in the jet, not taken into account here.
Uncertainties in the calculation of the physical parameters from
the characteristic wavelengths following Eqs.~(\ref{long}) and
(\ref{maxim}), as discussed in previous paragraphs, can be a
source of error in the determination of the parameters of the jet,
which, in turn, influence the long term stability properties.

Disruption of the jet in the simulation contradicts apparently the
fact that the jet is observed on much larger scales. It should be
noted however that the disruption point still propagates outward
at $v\sim 0.03\,c$, at the end of the simulations. This implies
that the simulation has not run long enough to reach a
quasi-steady stage. It is also possible that the disruption
observed is a transitory phase and that the simulation should have
run longer in order to allow the jet to move downstream. If a
stabilizing factor is needed in order to explain the jet in 3C273,
we suggest several possibilities: 1)~a thicker shear layer
(Birkinshaw \cite{bir91}, Hardee \& Hugues \cite{hah03}, Perucho
et al. \cite{pe+06}), 2)~ inclusion of the superluminal components
in the simulations, as faster jets are much more stable against
K-H instability (see Perucho et al. \cite{pe+04b}, Perucho et al.
\cite{pe+05}), 3)~a decreasing density atmosphere (Hardee
\cite{har82,har87}, Hardee et al. \cite{har05}, where it is shown,
in the case of the jet in 3C~120, that the expansion of the jet
provides a stabilizing influence), which must be the case as can
be derived from the outward dimming due to adiabatic expansion of
the observed jet in the parsec scale , 4)~a stabilizing
configuration of magnetic field (Rosen et al. \cite{ro+99}, Frank
et al. \cite{fra+96}, Jones et al. \cite{jon+97}, Ryu et al.
\cite{ryu+00}, Asada et al. \cite{asa+02}). The cumulative effect
of these factors would effectively make the jet more stable
already on the time scales probed by the simulation.

\subsection{Simulation 3C273-B}
\subsubsection{Initial setup}

In this simulation, the stationary flow is perturbed by precession
and periodical ejections of faster components. The initial
conditions are similar to those in the first simulation. The width
of the shear layer is $\sim 0.6\,R_j$ ($m=8$). The precession
frequency is derived from the observed $\sim 15$\,yr ($6\,R_j/c$
in the code units) periodicity of position angle variations
(Abraham et al. \cite{abr+96}). The frequency of ejections of
components is set by the reported $\sim 1$\,yr ($0.4\,R_j/c$)
periodicity in the optical light curve (Babadzhanyants and Belokon
\cite{bb93}). The duration of each ejection is estimated to be
$\sim 2$\,months ($0.066\,R_j/c$), set by the approximate
inspiralling time from an orbit at $\approx 6R_\mathrm{G}$ around
a $5.5\times 10^8\mathrm{M}_{\odot}$ black hole. The amplitude of
the precession is set by the true opening angle ($0.4^o$) of the
jet obtained by deprojecting the apparent opening of $1.5^o$. This
precession is included in transversal velocities at the injection
point as follows:
\begin{equation}
\overrightarrow{v_{\perp}}=A_0 v_z (\cos(\omega t),\sin(\omega t))
\end{equation}
where $\overrightarrow{v_{\perp}}$ and $v_z$ are the transversal
and axial components of the velocity, $t$ is time,
$A_0=6.83\,10^{-3}$ is the initial amplitude, and
$\omega\sim1.01\,c/R_j$ is the angular frequency calculated from
the precession period of $\sim 15\,\rm{yrs}$.

The fluid in the injected components is considered to have the
same density as the underlying flow and to be in pressure
equilibrium with it. Velocity of the fluid in components is taken
as constant, with the mean value of Lorentz factor
$\gamma_\mathrm{c}\simeq 5$ as reported in Abraham et al.
(\cite{abr+96}). The components are generated as shells of fluid
with a diameter of $0.5\,R_j$ ejected along the axis.

The numerical grid for this simulation covers $30\,R_j$ (axial)
times $6\,R_j$ times $6\,R_j$ (transversal), i.e.,
$24\,\rm{pc}\times 4.8\,\rm{pc}\times 4.8\,\rm{pc}$. The axial
dimension of the grid is related to the ejection of the
components. We take into account that the wavelength induced by
the precession of components, if they move ballistically, is
$\lambda_\mathrm{c} \sim P\,v_\mathrm{c}$, where $P$ is the
precession period and $v_\mathrm{c}$ is the injection velocity of
the fluid in the components. This gives $\lambda_\mathrm{c} \sim
6\,R_j$, and we have chosen the grid of $5\,\lambda_\mathrm{c}$ to
allow the wave to become apparent.

The resolution of the grid is 16 cells$/R_j$ in the transversal
direction and 32 cells$/R_j$ in the direction of the flow. An
extended grid is introduced in both transversal and axial
directions. In the radial directions, it has 36 cells reaching out
to $15\,R_j$ on each side of the jet (increasing the cell size by
$7.7\%$ from one cell to the next). In the axial direction the
extended grid has 192 cells, reaching up to $45\,R_j$. This
simulation has lasted for a time of $70\,R_j/c$ (i.e., more than
two light crossing times of the grid).

\subsubsection{Results}
\begin{figure*}[!t]
\centerline{
  \includegraphics[width=0.8\textwidth]{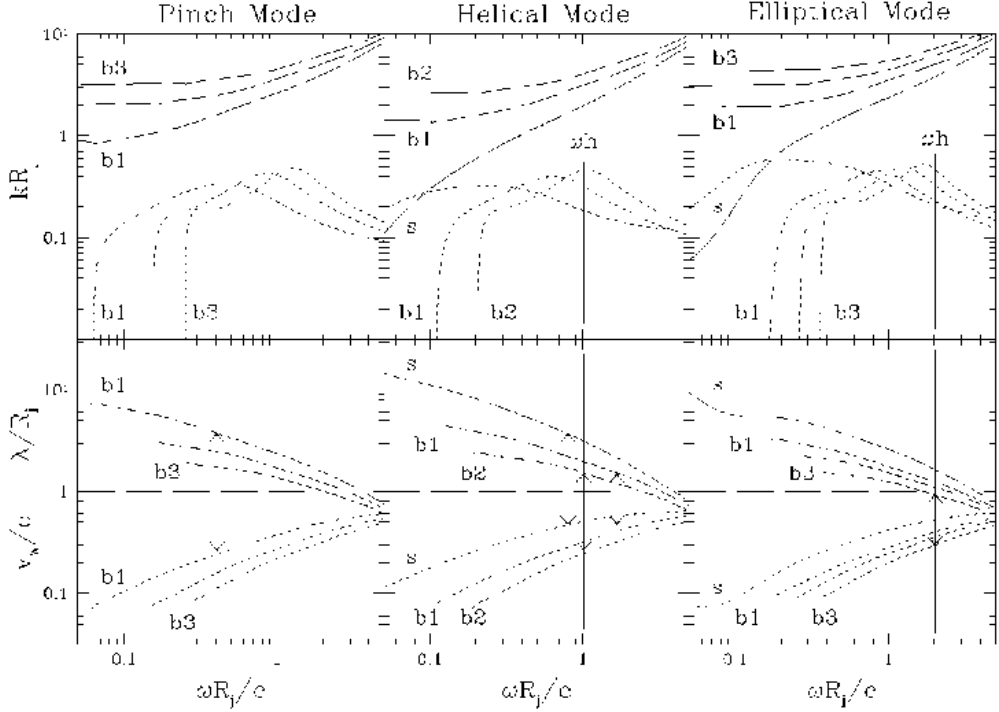}}
\caption{Solution of the stability problem for the parameters
given in Table \ref{tab:3c2732} with the characteristic
frequencies for simulation 3C273-B. Frequency $\omega_h=1.01c/R_j$
corresponds to that excited in the simulation 3C273-B. Arrows
connecting the wavelength and wave speed plots indicate identified
modes in the simulation 3C273-B (see text).}\label{fig:drs-B}
\end{figure*}

\begin{figure}[!t]
   \centering
   \resizebox{0.8\hsize}{!}{\includegraphics{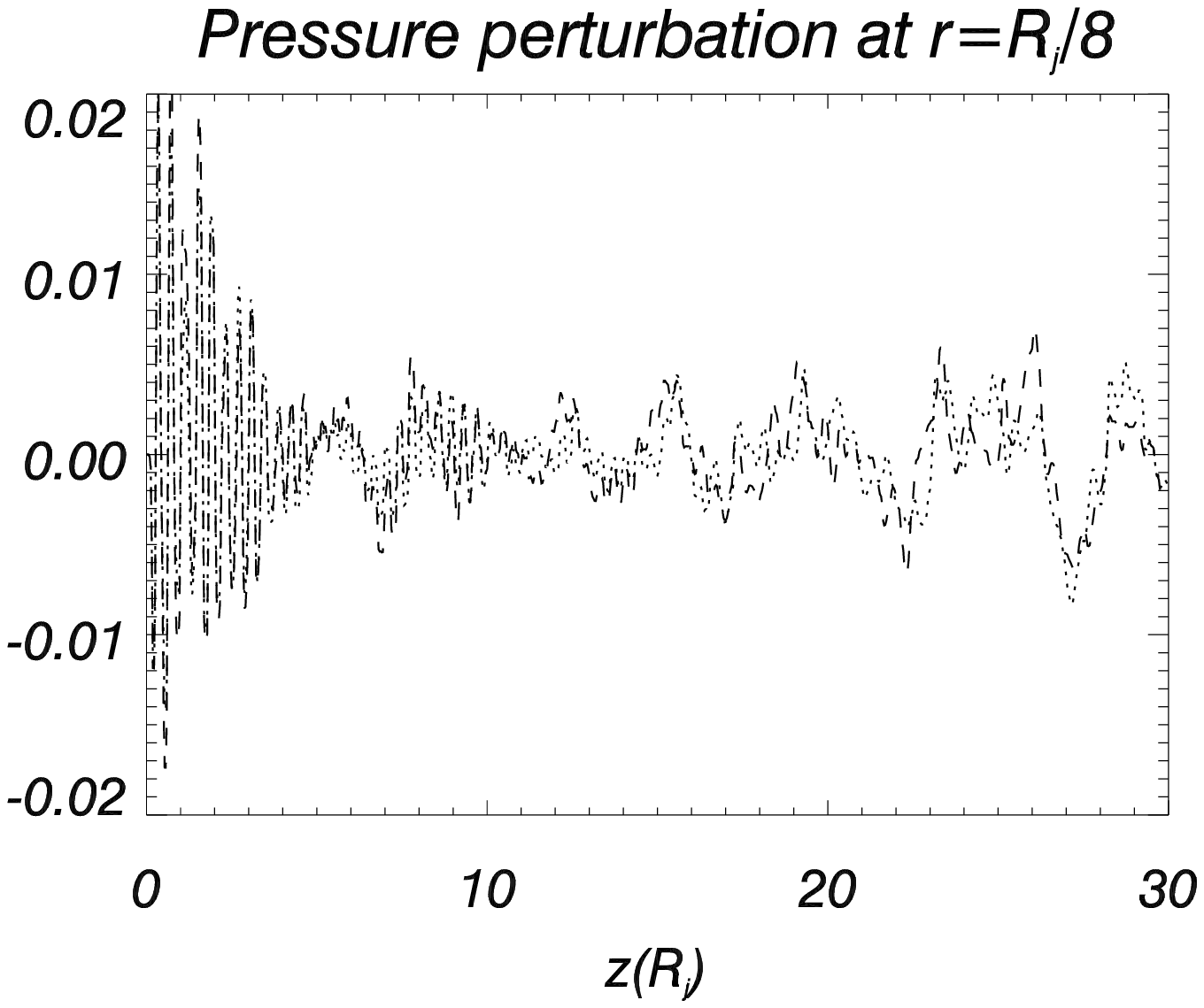} \includegraphics{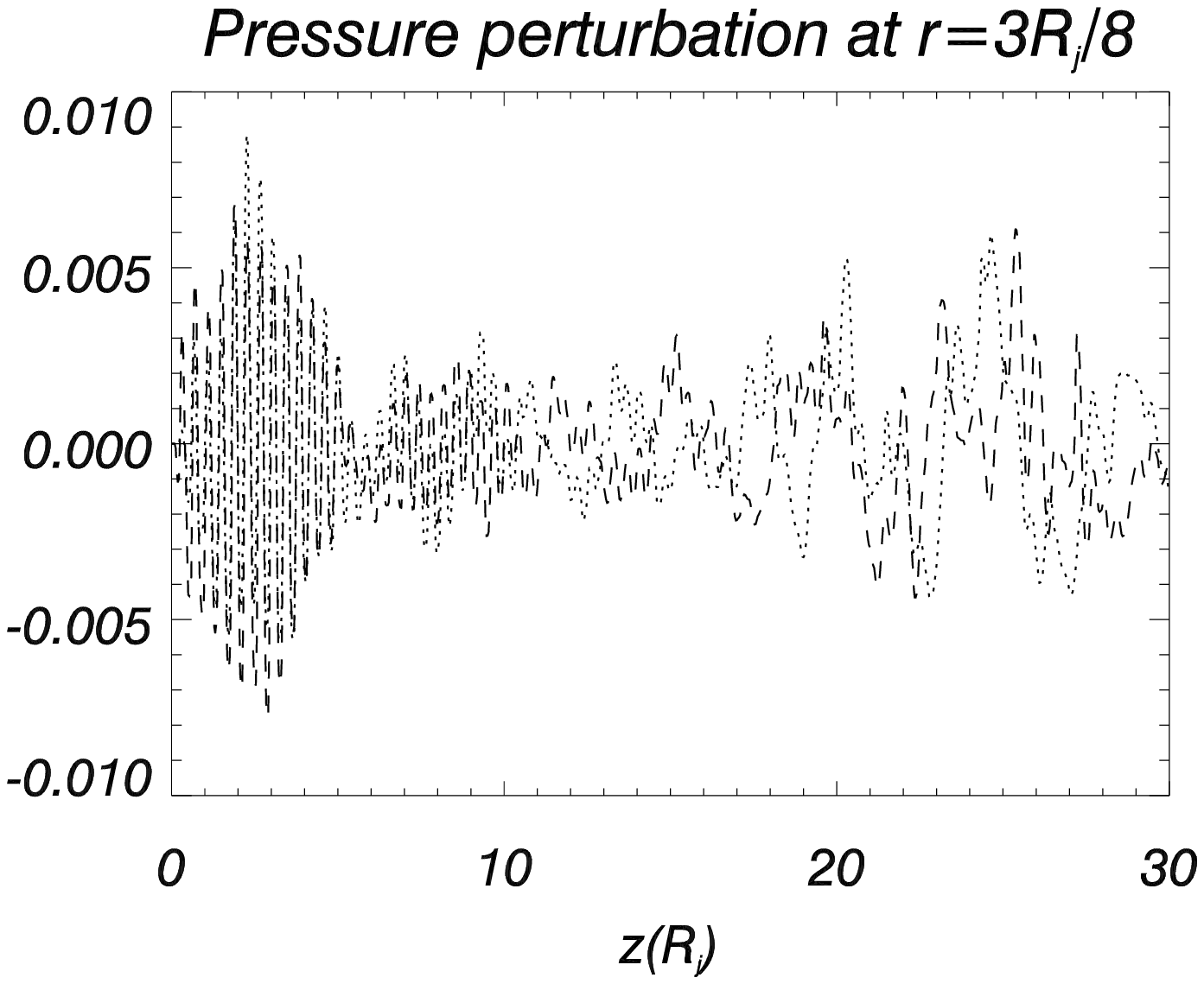}}
   \quad \resizebox{0.8\hsize}{!}{\includegraphics{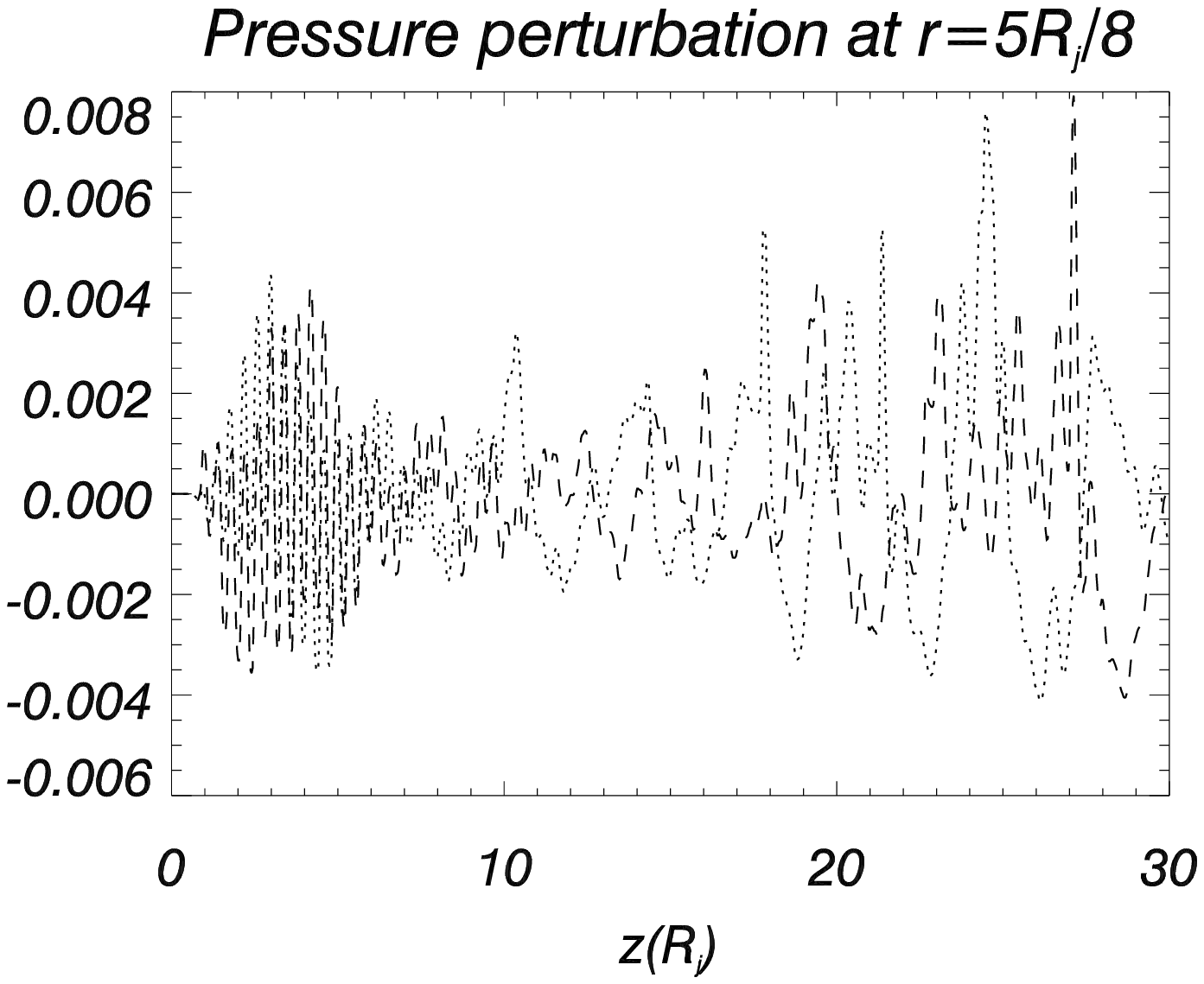} \includegraphics{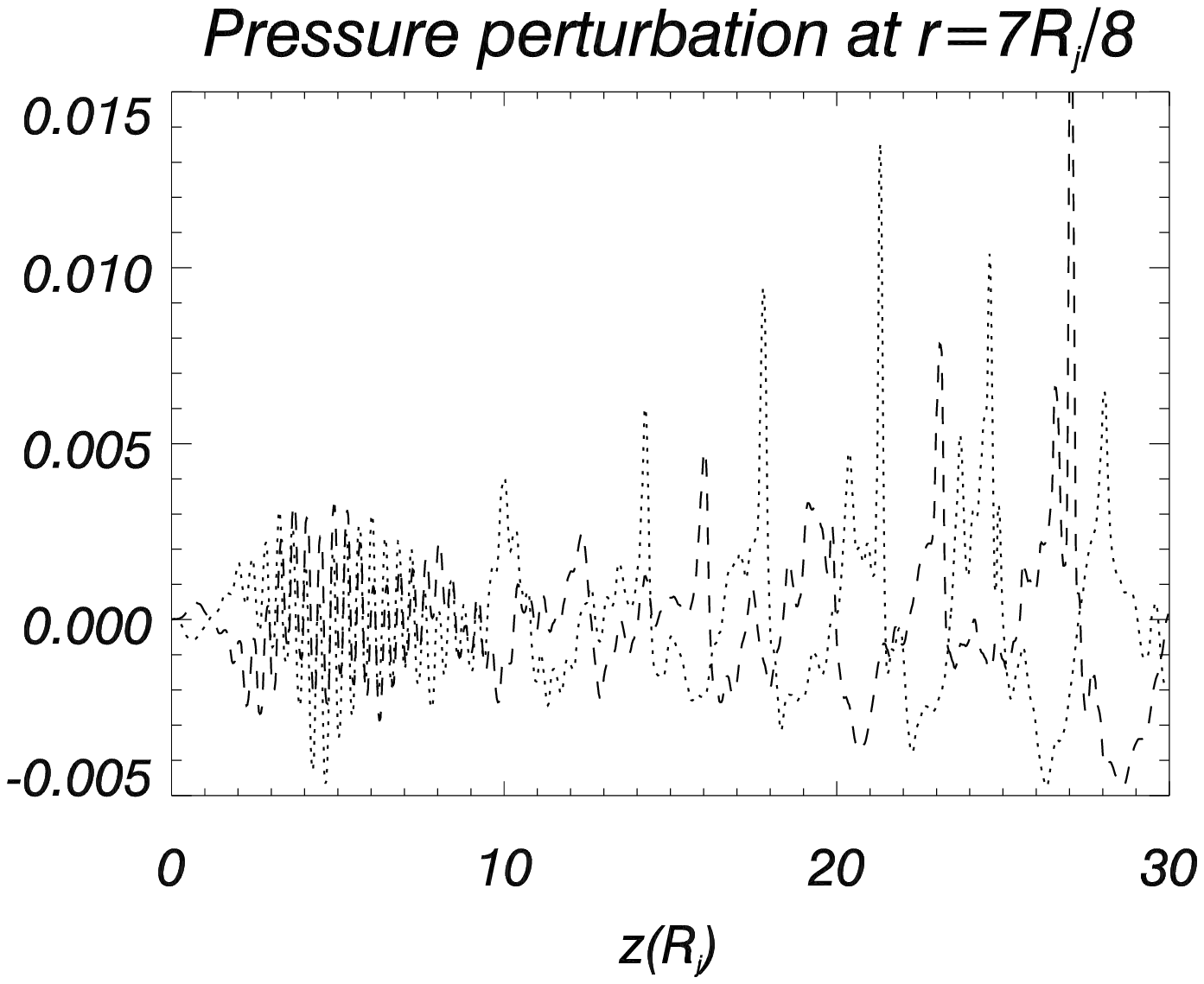}}
\caption{Longitudinal cuts of pressure perturbation at different
radii ($R_j/8$ top left, $3R_j/8$ top right, $5R_j/8$ bottom left,
$7R_j/8$ bottom right) and in symmetric positions with respect to
the jet axis at $t=40\,R_j/c$. Short wavelength
($\lambda^{sim}_1=0.4\, R_j$ and $\lambda^{sim}_3=0.9\, R_j$)
structures are observed close to the jet axis. The amplitude of
the first structures decrease downstream as a consequence of the
drift of components to outer radial positions due to the
precessional motion and their interaction with the underlying
flow. Also at positions close to the axis, we find a structure
with wavelength $\lambda^{sim}_3=1.5\, R_j$. Short wavelength
symmetric ($\lambda^{sim}_1=0.4\, R_j$) and longer wavelength
antisymmetric ($\lambda^{sim}_2=4\, R_j$) structures are also
observed at outer radial positions. The longer structure dominates
close to the injection, before the components expand and/or drift
from the axis to the jet boundaries, when short scale structures
can be observed to be modulated by the longer one (from $z\sim
4\,R_j$). After this, at $z=9\,R_j$, the antisymmetric longer
structure grows in amplitude, with some spikes that can be
associated with the drift of the components due to their
precessional motion.} \label{pressb1}
\end{figure}

Fig.~\ref{fig:drs-B} shows the solutions of the linear stability
problem for the jet parameters given in Table \ref{tab:3c2732} and
indicates the characteristic wavelengths arising un this
simulation. Fig.~\ref{pressb1} shows longitudinal cuts of pressure
perturbation at different jet radii at time $40\,R_j/c$.
Inspection of the longitudinal cuts indicates that, close to the
injection point and to the jet axis, a symmetric, short wavelength
perturbation generated by the fast components dominates the
structure of the flow. Its wavelength is $\lambda^{sim}_1=0.4\,
R_j$, and it is clearly related to the ejection period of
components ($0.4\, R_j/c$). After $3-4\, R_j$, the presence of the
fast components is also evident in the jet boundary (at
$r=7/8R_j$). These high-frequency and short-wavelength structures
are damped at $z>10\,R_j$, as expected to occur for perturbations
with wavelengths smaller than the shear layer width. Close to the
jet boundary, the most pronounced structure is the typical
antisymmetric pattern of helical motion induced by the precession
($\lambda^{sim}_2\sim 3.7\, R_j$). This structure is driven by the
precession of the injected components and it couples to the
helical surface mode at a frequency of $0.8\, c/R_j$. A pinch mode
structure with a wavelength $\lambda^{sim}_2=3.7\,R_j$ is also
observed at $r=R_j/8$, coinciding with the maximum growth of the
first body mode (see Fig.~\ref{fig:drs-B} and
Table~\ref{tab:drsh}).

\begin{figure}[!t]
   \centering
   \resizebox{0.8\hsize}{!}{\includegraphics{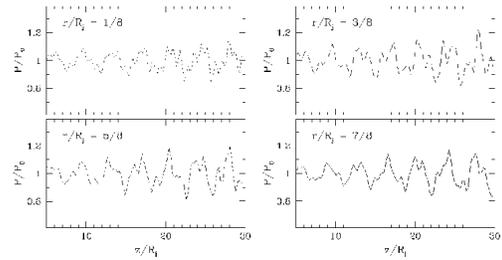}}
   \caption{Theoretical computation of the pressure perturbations produced
by the combination of modes identified in the simulation.
Different panels indicate cuts at different distances from the
axis in the X-Z plane. The wavelengths and wave speeds of the
modes contributing to this perturbations are marked by arrows in
Fig.~\ref{fig:drs-B}. These are the first pinch body mode at
$\lambda=3.7\,R_j$, the helical surface mode at
$\lambda=3.7\,R_j$, helical first and second body modes at
$\lambda=1.5\,R_j$,  and the elliptical third body mode at
$\lambda=0.9\,R_j$. We have used a linearly growing amplitude for
the modes. This perturbations can be compared with the structures
identified in Fig.~\ref{pressb1}. Main differences between this
plot and Fig.~\ref{pressb1} are due to nonlinearities introduced
by injected components, mainly close to the injection on the jet
axis and at outer radii farther downstream.}
     \label{harcutsb}
\end{figure}

Farther downstream and close to the jet center, there is a
$\lambda^{sim}_3=0.9\,R_j$ wavelength on top of the longer pinch
mode ($\lambda^{sim}_2$). This short wavelength coincides with the
maximum growth rate of the elliptical third body mode at the
helical driving frequency ($1.01\, c/R_j$) (see
Fig.~\ref{fig:drs-B} and Table~\ref{tab:drsh}). We remind the
reader that this is frequency correspond to a 360$^\circ$ turn, as
explained in the caption of Fig.~\ref{fig:drs}. The radial
structure of this mode in the simulation is also coincident with
the theoretical structure of the third body elliptical mode, with
the maximum amplitude at $r=R_j/4$ and decreasing amplitude at
larger radii. The helical driving frequency falls very close to
the maximum growth of the second body helical mode with wavelength
$\lambda^{sim}_4=1.5\,R_j$, observed in Fig.~\ref{pressb1}.
However, the radial structure found in the simulation is somehow a
mixture of the second body mode with that of the first body mode.
We expect the second body mode to have an amplitude maximum at
$r=1/8R_j$, but we find that this maximum occurs at $r=3/8R_j$.
The first body mode at this wavelength develops an amplitude
maximum at $r=3/8R_j$, which is coincident with the one found in
the simulation, although the amplitude at $r=1/8R_j$ found in the
simulation is too large for this mode. We interpret this as the
second body mode being triggered at the driving frequency, which
in turn excites the first body mode at the same wavelength. Both
modes seem to be triggered out of phase and interfere
destructively in the inner jet, but the first body mode dominates
in the mid jet region and both decline in amplitude towards the
jet surface. Comparison of smaller and larger radius plots of
pressure perturbation in Fig.~\ref{pressb1} shows large positive
offsets observed at larger radii. This indicates possible drifting
of components (shocks) to outer radii of the jet as they follow
the helical path given by the surface mode (similar to what has
been reported by LZ01). This feature would produce enhanced
emission regions at the positions of the helix in which the flow
moves in a direction closer to the line of sight. Using all of the
modes identified in the simulation we have computed a theoretical
representation of the pressure perturbation in the jet.
Fig.~\ref{harcutsb} shows the resulting axial cuts of the pressure
perturbation and indicates that the theoretical calculation
generates most of the structures found in the simulation, except
those which are intrinsically nonlinear (for instance, the
injected components at the jet inlet and those structures at outer
jet radii farther downstream).

%----------------------------------------------------------- S_vib

%                                                Two column figure
%----------------------------------------------------------- S_vib
%
%                                                Two column figure
%----------------------------------------------------------- S_vib

\begin{figure*}
\centerline{
\psfig{file=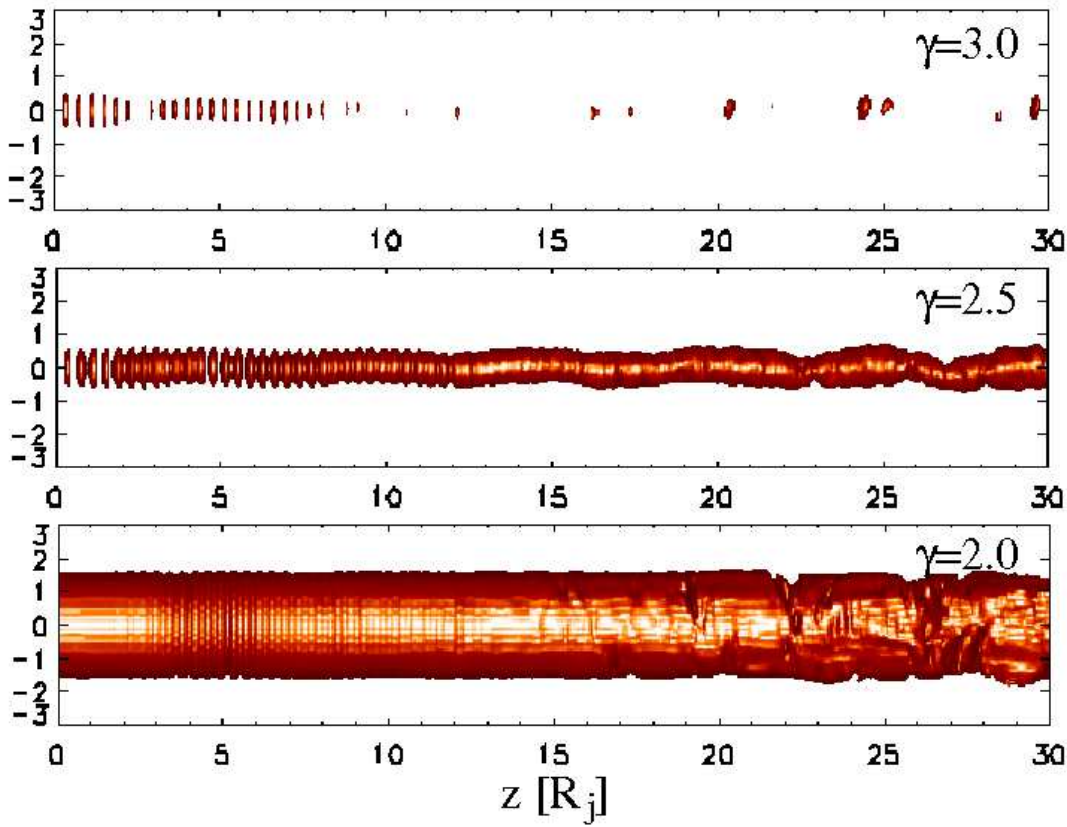,width=0.8\textwidth,angle=0,clip=}}
\caption{Surface plots for three different values of Lorentz
factor: Upper panel, $\gamma=3.0$; middle panel, $\gamma=2.5$;
lower panel, $\gamma=2.0$. The precession motion, coupled to a
helical mode, is more apparent on the slower, underlying flow,
moving at $\gamma=2.0-2.5$, as shown in LZ01. The surface plot for
$\gamma=3.0$ illustrates the appearance of the jet at higher
frequencies and smaller viewing angles, where the emission is
dominated by a more relativistic plasma.} \label{lof3}
\end{figure*}

The complexity of the structure is further illustrated by the
surface plot of the flow Lorentz factor shown in Fig.~\ref{lof3}.
Although in this figure we only see patterns in the fluid,
comparison of the wavelengths seen here and in Fig.~\ref{pressb1}
allows us to identify the wavelengths derived from the fluid
patterns in Fig.~\ref{lof3} and from the frequency of injection of
components. The middle panel of Fig.~\ref{lof3} (Lorentz factor
$\gamma=2.5$) indicates that the periodicity induced by individual
jet components dominates the structure up to distances
$z\sim10-15\,R_j$, but farther downstream the components expand
longitudinally and start to interact with each other, generating a
semi-continuous structure that is dominated by the helical motion
induced by the precession. The top panle of Fig.~\ref{lof3}
(Lorentz factor $\gamma=3.0$) indicates that the distinct regions
of the flow moving at higher speed disappear downstream. This can
be explained by the deceleration of the fluid in the components.
The deceleration can be caused either by the interaction with the
background flow, or by a radial and longitudinal expansion. These
results are in agreement with results by Lobanov and Zensus
(\cite{lz99}), and Lobanov and Roland (\cite{lr01}), suggesting
that shocks dominate the jet structure close to the nucleus, and
that fluid instabilities become important farther downstream. On
the other hand, it is not clear at the moment whether stronger
fluid components (i.e., faster and denser) could survive longer in
the jet. We finally note that the precession wavelength obtained
from the simulations is smaller than the one calculated
theoretically from the advance speed of components
($\lambda^{sim}_2=4\,R_j$ versus $\lambda_c\sim 6\,R_j$), which
maybe taken as an indication of non-ballistic motion of the
components.

\begin{table*}
\caption{First column gives the observed wavelengths in the
simulation, and the last three columns give this wavelengths as
observed depending on the wave speed. We have used, for the wave
speed, some of those obtained from Fig.~\ref{fig:drs-B} for the
identified modes (second to fourth columns), the underlying flow
speed (fifth column) and the maximum speed from the components
(sixth column).} \label{tab:t0b} \centering
\begin{tabular}{c|ccccc}
\hline \hline $\lambda^{sim}$ [$R_j$]&
$\lambda^{obs}_{v_w=0.25\,c}$[mas]
&$\lambda^{obs}_{v_w=0.4\,c}$[mas]
&$\lambda^{obs}_{v_w=0.5\,c}$[mas]
&$\lambda^{obs}_{v_w=0.88\,c}$[mas] &
$\lambda^{obs}_{v_w=0.94\,c}$[mas]\\
\hline
0.4 &0.045&0.05 &0.07&0.23&0.37\\
0.9 &0.1&0.12 &0.15&0.51&0.83 \\
1.5 &0.17& 0.2&0.25&0.85&1.4\\
4 &0.45&0.53&0.67&2.27&3.7\\
\hline
\end{tabular}
\end{table*}

In Table~\ref{tab:t0b} we list possible observed wavelengths
corresponding to the two main wavelengths identified from the
simulation. It is clear that the observed wavelength of precession
($18\,\rm{mas}$) cannot be recovered even with extremely fast
components ($v_w=c$ gives $\lambda^{obs}=10 mas$) with the adopted
viewing angle of $15^\circ$. Moreover, if we consider the mean
apparent proper motion of $0.93\,h^{-1}\rm{mas/yr}$ (Abraham et
al. \cite{abr+96}), the apparent speed is $\beta_{app}\sim 10$.
This speed cannot be reconciled with a viewing angle of
$15^\circ$, as the resulting intrinsic speed is larger than 1, and
it would require a Lorentz factor of $\gamma=10$ if the viewing
angle is reduced to $10^\circ$. Recent work by Jorstad et al.
(\cite{jor+05}) indicates that the jet in 3C273 can have a viewing
angle as small as $6^\circ$ and component Lorentz factors of
$\gamma \sim 10.6$. These numbers would transform
$\lambda^{sim}_2=4\,R_j$ into an observed wavelength $14.1\,
\rm{mas}$. This wavelength is in fair agreement with the $18\,
\rm{mas}$ mode assigned to precession in LZ01. In summary, either
the precession period should be longer, or the viewing angle
should be smaller than $15^\circ$ and component Lorentz factors
higher than $\gamma=3$, in order to reconcile the 18 mas structure
with precession.

Another question addressed by this simulation is whether the
periodic injection of fast components could generate smaller
structures observed by LZ01 (the $2\, \rm{mas}$ and $4\, \rm{mas}$
modes), where these wavelengths are identified with the elliptical
modes of Kelvin-Helmholtz instability. In our simulation, the fast
components generate mainly pinching modes, although this is simply
due to their symmetric nature. Again, we find that the structures
generated in the simulation are small compared to those observed.
However, in the simulation, the fluid in the components moves with
a smaller speed than the injection one ($\gamma=5$, see
Fig.~\ref{lof3}) and therefore, smaller than that found from the
observations ($\gamma\sim5-10$). We relate this fact to the
possible slowing of the components themselves as they propagate
downstream. As pointed out before, a possible cause for this could
be that the simulated fluid injected in components has the same
density as the background flow, whereas if it was denser (as could
be expected as generated from strong accretion activity) or
propagating in a decreasing density atmosphere, this fluid would
have larger inertia and would generate faster components. We find
that, in order to produce a 4 mas structure, we need the Lorentz
factor of components to be $\gamma\sim 30$, whereas $\gamma\sim15$
is required to explain the 2 mas wavelength, if we keep the 1 yr
period. Adopting the longest measured period of the ejections of
1.7 yrs (Abraham and Romero \cite{ar99}), these values would be
reduced to $\gamma\sim17$ and $\gamma\sim8.5$, respectively. The
same authors gave a periodicity in the injection Lorentz factor of
about 4 yrs; if we consider this period as the generator of short
modes, $\gamma\sim7.5$ and $\gamma\sim4$ could explain those
structures. The latter values agree well with the Lorentz factors
inferred from the observed kinematics of the jet. This means that
the 2 and 4 mas wavelengths should be associated only with the
strongest and fastest ejections occurring roughly once every 4
yrs. With the jet parameters given by Jorstad et al.
(\cite{jor+05}), the inferred wavelength for the shortest
symmetric structure ($0.4\,R_j$) is $1.4\, \rm{mas}$, which is
within a factor of 3 from the 4 mas wavelength identified in LZ01
with the elliptical first body mode. The symmetric or pinching
nature of the perturbations induced by such components could not
explain the double helix structure in the jet, but we observe in
the simulation that elliptical modes can be triggered by the
presence of pinching and helical perturbations.

\section{Conclusions}
\label{sect4}

We have performed two numerical RHD simulations with different
initial setups in order to study the physical processes generating
the observed structures in the parsec-scale radio jet in the
quasar 3C\,273. In the simulation 3C273-A, we have included a
general set of helical and elliptical perturbations in a long jet
with the basic physical parameters adopted from LZ01. In the
simulation 3C273-B, we have used a shorter jet with the same
physical parameters and have included precession and injection of
fast components. The simulation 3C273-A was aimed to generate
structures with wavelengths similar to those measured by LZ01 from
the growth of Kelvin-Helmholtz perturbations. The simulation
3C273-B was designed to check if by combining the ejection of
\emph{superluminal} components and jet precession, with the
periodicities reported in Babadzhanyants and Belokon (\cite{bb93})
and Abraham et al. (\cite{abr+96}), the same structures could be
generated.

We find that the structures generated in simulation 3C273-A are of
the same order in size as those observed, if the relativistic
propagation effects of the waves are taken into account. We
observe in the solution of the stability problem that the
instability modes found in the simulation propagate at mildly
relativistic speeds. These wave speeds differ from those derived
from the approximations used in LZ01. This can be due to the
uncertainties introduced by the approximations to the
characteristic wavelengths in the interpretation of the
observations in LZ01. However, we show that wavelengths similar to
the observed ones are found for the wave speed given by the
solution of the linear problem, although the modes fitted in LZ01
and those used here for the same wavelengths are not coincident.
The solutions of the stability problem applied to the adopted wave
speeds (0.23 $c$) and line of sight ($15^\circ$) show that any
body modes present in the jet should be much shorter than those
fitted in LZ01. It should be noted that these differences do not
rule out the presence of Kelvin-Helmholtz instability in
parsec-scale jets. Despite difficulties in the mode
identifications, the structures generated in the simulation are
similar to those observed by LZ01.

Regarding the long-term stability of the flow, we note that the
jet in the simulation 3C273-A is disrupted at $\sim 170\, \rm{pc}$
from the inlet, contrary to the observations tracing the jet in
3C273 up to $60\, \rm{kpc}$ away from the source. The reasons for
this difference may be found in the conjunction of several
factors. 1)~The numerical simulation does not run long enough to
reach a fully steady-state regime. The disruption point moves
downstream along the simulation, which could imply that the
disruption is a transitory phase. 2)~Magnetic fields have not been
taken into account neither in the linear analysis, nor in the
numerical simulation - and it should be noted that the magnetic
fields may be dynamically important at parsec scales (Rosen et al.
\cite{ro+99}, Frank et al. \cite{fra+96}, Jones et al.
\cite{jon+97}, Ryu et al. \cite{ryu+00}, Asada et al.
\cite{asa+02}). 3)~We only simulate the underlying flow, without
considering the faster and possibly denser fluid in the
superluminal components. 4) Inaccuracies in the linear analysis
approximations can lead to large uncertainties in physical
parameters derived. 5)~Differential rotation of the jet, shear
layer thickness (Birkinshaw \cite{bir91}, Hardee \& Hugues
\cite{hah03}, Perucho et al. \cite{pe+06}), and a decreasing
density external medium could also play an important role
(implying jet expansion; see Hardee \cite{har82,har87}, and Hardee
et al. \cite{har05}). 6)~Arbitrary initial amplitudes of
perturbations were chosen for the simulation, so we could have
included too large perturbations. A combination of these factors
could well change the picture of the evolution of the jet in terms
of its stability properties. The effects of the rotation and
magnetic fields on the stability of jets remain unclear, since no
systematic numerical study has been performed up to now.

In the simulation 3C273-B, we studied the effect of precession on
the jet evolution and investigated the possibility that the short
wavelength structures found in LZ01 were not due to K-H
instabilities but due to the periodicities induced in the flow by
the ejection of components. We demonstrate that such non-linear
features as superluminal components generate linear structures in
the form of Kelvin-Helmholtz instabilities which can be analyzed
in the framework of linear perturbation analysis. One of the main
conclusions that can be derived from this simulation is that
helical twists can be excited by periodic injections if there is
some induced helicity in the system. This helicity is induced in
our simulation by the helical perturbation frequency, but in real
jets this helicity could be induced by helical jet magnetic fields
and/or by jet rotation. We have also shown that, in order to
explain the observed $18\,\rm{mas}$ wavelength in terms of
precession, either longer driving periodicities than the
$15\,\rm{yr}$ suggested by Abraham \& Romero (\cite{ar99}) would
be needed, or this wavelength must be induced by very fast
components observed in a jet moving at a viewing angle
$\theta<15^\circ$. The fast components could only generate the
shorter wavelengths given in LZ01 ($\lambda=2$ mas and $\lambda=4$
mas) if a proper combination of the velocities and injection
periodicities is used. Altogether, we find that inclusion of
faster components and precession with the $15\,\rm{yr}$
periodicity does not explain well the observed wavelengths and
periodicities. This gives more weight to the general conclusion
about K-H instability acting prominently in the flow.

In the future, numerical simulations of this kind may be used to
constrain the basic parameters of the flow such as the viewing
angle and the component speed. Inclusion of magnetic fields,
differential rotation and the effects of an atmosphere with a
decreasing density could help reconciling better the simulations
with the observed structures. In this way, for example, an
increase of the jet radius due to decreasing external density
could cause a downstream increase of wavelengths of K-H
instability modes (Hardee et al. \cite{har05}). This remains to be
seen with future, full-fledged RMHD simulations of the
relativistic jet in 3C~273. The scope of the present work could be
expanded to other sources, and applied to prominent jets for which
the transversal structure may be resolved, such as 3C~120,
extending the work done by Hardee et al. (\cite{har05}) by
performing numerical simulations.

\acknowledgements{Calculations were performed on the SGI Altix
3000 computer "CERCA" at the Servei d'Inform\`atica de la
Universitat de Val\`encia. This work was supported in part by the
Spanish Direcci\'on General de Ense\~nanza Superior under grants
AYA-2001-3490-C02 and AYA2004-08067-C03-01 and Conselleria
d'Empresa, Universitat i Ciencia de la Generalitat Valenciana
under project GV2005/244. M.P. benefited from a predoctoral
fellowship of the Universitat de Val\`encia ({\it V Segles}
program) and a postdoctoral fellowship in the Max-Planck-Institut
f\"ur Radioastronomie in Bonn. P. Hardee acknowledges support by
NSF award AST-0506666 and through NSSTC/NASA cooperative agreement
NCC8-256 to The University of Alabama.}

\end{document}